\documentclass[journal]{IEEEtran}
\usepackage[pdftex]{graphicx}
\graphicspath{{../pdf/}{../jpeg/}}
\DeclareGraphicsExtensions{.pdf,.jpeg,.png}
\usepackage[cmex10]{amsmath}
\usepackage{url}
\usepackage[table,xcdraw]{xcolor}
\usepackage{graphicx,color}
\usepackage{graphicx}
\usepackage{amssymb,color,amsfonts,mathrsfs,balance,mathtools,xfrac,nccmath, siunitx}
\allowdisplaybreaks
\usepackage{subcaption}
\usepackage{cite}
\usepackage{blindtext}
\usepackage{tcolorbox}
\usepackage{bbm}
\usepackage{xcolor}
\usepackage{lipsum}
\usepackage[font={small}]{caption}
\usepackage{algorithm}
\usepackage{algorithmic}
\usepackage{array} 
\usepackage{makecell}
\usepackage{tabularx}
\definecolor{Gray}{gray}{0.9}
\definecolor{LightGray}{gray}{0.9}

\newtheorem{remark}{Remark}

\newcommand{\q}[1]{{\bf #1}}

\newcommand{\qh}{\mathbf{h}}
\newcommand{\qa}{\mathbf{a}}
\newcommand{\qs}{\mathbf{s}}
\newcommand{\qg}{\mathbf{g}}
\newcommand{\qr}{\mathbf{r}}
\newcommand{\qu}{\mathbf{u}}
\newcommand{\qy}{\mathbf{y}}
\newcommand{\qw}{\mathbf{w}}

\newcommand{\qz}{\mathbf{z}}

\newcommand{\qF}{\mathbf{F}}

\newcommand{\qW}{\mathbf{W}}
\newcommand{\qI}{\mathbf{I}}

\newcommand{\qQ}{\mathbf{Q}}

\usepackage{bm}

\newcommand{\qtheta}{\bm{\theta}}

\newcommand{\qpsi}{\boldsymbol{\psi}}

\newcommand{\tr}{\text{Tr}}

\newcommand{\Pth}{P_{\text{th}}}

\newcommand{\bbc}{\text{BiBC}\xspace}

\newcolumntype{L}[1]{>{\raggedright\arraybackslash}p{#1}}
\newcolumntype{C}[1]{>{\centering\arraybackslash}p{#1}}
\newcolumntype{R}[1]{>{\raggedleft\arraybackslash}p{#1}}

\begin{document}
	\bstctlcite{IEEEexample:BSTcontrol}

	\title{Refined-Deep Reinforcement Learning for MIMO Bistatic Backscatter Resource Allocation}

	\author{Shayan Zargari, Diluka Galappaththige, \IEEEmembership{Member, IEEE},  and Chintha Tellambura, \IEEEmembership{Fellow, IEEE} 
		\thanks{S. Zargari, D. Galappaththige, and C. Tellambura with the Department of Electrical and Computer Engineering, University of Alberta, Edmonton, AB, T6G 1H9, Canada (e-mail: \{zargari, diluka.lg, ct4\}@ualberta.ca).} \vspace{-5mm}}

	\maketitle 
	
	\begin{abstract} 
		Bistatic backscatter communication facilitates ubiquitous, massive connectivity of passive tags for future Internet-of-Things (IoT) networks. The tags communicate with readers by reflecting carrier emitter (CE) signals. This work addresses the joint design of the transmit/receive beamformers at the CE/reader and the reflection coefficient of the tag. A throughput maximization problem is formulated to satisfy the tag requirements. A joint design is developed through a series of trial-and-error interactions within the environment, driven by a predefined reward system in a continuous state and action context. By leveraging recent advances in deep reinforcement learning (DRL), the underlying optimization problem is addressed. Capitalizing on deep deterministic policy gradient (DDPG) and soft actor-critic (SAC), we proposed two new algorithms, namely refined-DDPG for MIMO \bbc (RDMB) and refined-SAC for MIMO \bbc (RSMB). Simulation results show that the proposed algorithms can effectively learn from the environment and progressively improve their performance. They achieve results comparable to two leading benchmarks: alternating optimization (AO) and several DRL methods, including deep Q-network (DQN), double deep Q-network (DDQN), and dueling DQN (DuelDQN). For a system with twelve antennas, RSMB leads with a \qty{26.76}{\percent} gain over DQN, followed by AO and RSMB at \qty{23.02}{\percent} and \qty{19.16}{\percent}, respectively. 
	\end{abstract}
	
	
	\IEEEpeerreviewmaketitle
	
	\section{Introduction}
	Bistatic backscatter communication (\bbc) networks, which may enable energy harvesting (EH)-based Internet-of-Things (IoT) networks, are of significant interest as the third-generation partnership project (3GPP) has initiated a new study item \cite{3GPPAmbIoT23}. Applications of \bbc-enabled IoT networks include logistics, inventory management, warehousing, manufacturing, energy industry, healthcare, agriculture, aerospace and defense, farming, retail, sports, and many more \cite{Diluka2022, Rezaei2023Coding, Niu2019overview}. 
	
	A typical \bbc network consists of a dedicated carrier emitter (CE), a reader, and a backscatter device(s) or tag(s) \cite{Diluka2022, Rezaei2023Coding, Niu2019overview}. The key idea is that tags are passive devices that reflect radio frequency (RF) signals from the CE. The reflections enable tags to communicate with the reader  \cite{Zargari10353962}. \bbc with a standalone RF source located separately from the reader offers lower round-trip Pathloss, predictability, reduced interference, system control, and knowledge of source signal parameters over other backscatter configurations\footnote{Backscatter fundamentals and configurations, i.e., monostatic, bistatic, and ambient, can be found in \cite{Diluka2022, Rezaei2023Coding, Niu2019overview,Fatemeh_Rezaei}.}. Consequently, this study employs a \bbc system.

	\subsection{Related work}
	In \bbc systems, CE beamforming, tag reflection coefficients, and receive combiner at the reader must all be optimized to improve communication performance. The optimization is subject to the constraints on transmit power at the CE, minimum activation power targets for the tags, normalized combiner vectors, and unit-interval constraints for the reflection coefficients. Consequently, this is a non-convex and  NP-hard problem.  
	
	Consequently, related optimization studies can be classified into two categories: (1) classical methods without using machine learning (ML)   \cite{Sacarelo2021, Jia2022, Zhao2024, Diluka2024} and (2) ML-based methods  \cite{Anh2019, Anh2021, Zhang2019, Liu2021} methods. 
	
	Category one often uses alternating optimization (AO) \cite{bezdek2003} as its bedrock. AO is particularly effective for non-convex problems that can be decomposed into smaller, more manageable subproblems. Each subproblem is typically easier to solve than the overall problem and may be suitable for convex methods, closed-form solutions, or other optimization techniques. Applied to the current problem, AO can be used to iteratively optimize CE beamforming, tag reflection coefficients, and reader combiner until convergence is achieved.
	For example,  \cite{Sacarelo2021} uses successive convex approximation (SCA) to optimize CE beamforming, tag reflection coefficients, and reader combiners in a non-orthogonal multiple access (NOMA)-assisted \bbc system. References \cite{Jia2022, Zhao2024} study an intelligent reflecting surface (IRS)-assisted \bbc system to minimize the CE transmit power. In \cite{Jia2022}, two algorithms, i.e., minorization-maximization and AO, are proposed for single-tag and multi-tag scenarios, respectively, utilizing successive refinement and semidefinite relaxation (SDR) approaches to improve CE beamforming, tag reflection coefficients, reader combiners, and IRS phase-shifts. In \cite{Zhao2024}, an AO approach is presented for a single tag and a single-antenna reader system to construct CE transmit beamforming, tag reflection coefficient, and IRS phase shifts. Reference \cite{Diluka2024} offers an algorithm for designing beamforming at distributed CEs, tag reflection coefficients, and reader reception combiners utilizing AO, fractional programming (FP), and Rayleigh quotient techniques to maximize the tag sum rate while proposing a channel estimation protocol.
	
	However, such methods may converge to a local rather than a global optimum, especially in this non-convex problem. The process can be slow if the number of variables is large or if each subproblem is still computationally intensive. They also struggle with network complexity  \cite{Chowdhury2020}. These issues lead to Category two. 
	
	Deep reinforcement learning (DRL), a fusion of reinforcement learning (RL) and deep learning (DL) has emerged as a remedy \cite{Anh2019, Anh2021, Zhang2019, Liu2021}. Nevertheless, previous DRL works \cite{Anh2019, Anh2021, Zhang2019, Liu2021} only examine ambient backscatter networks and not \bbc. Moreover,  these studies focus on single antenna RF sources and single tag scenarios, except for \cite{Anh2021}. Although \cite{Anh2021} supports multiple tags, it is limited to a single-antenna source and single-antenna reader. 
	
	Reference \cite{Anh2019} employs a double deep $Q$-network (DDQN) to allow the reader to learn the best policy based on primary channel utilization for coordinating the transmission of multiple ambient backscatter tags, including backscattering time, energy harvesting time, and active transmission time. In \cite{Anh2021}, a deep deterministic policy gradient (DDPG) algorithm is used to optimize the backscatter relaying policies, i.e., reflection coefficients, to maximize network throughput in a backscatter-assisted relaying network. Reference \cite{Zhang2019} presents two algorithms: (1) constellation learning with labeled signals and (2)  constellation learning with labeled and unlabeled signals, using ML-based modulation-constrained expectation maximization algorithms. Study \cite{Liu2021} develops a deep transfer learning (DTL) detection framework that includes offline learning, transfer learning, and online detection of tag signals. It implicitly extracts channel features and recovers tag symbols, eliminating the need for channel estimation.
	
	DRL has an innate ability to intelligently and adaptively navigate complex optimization landscapes \cite{Li2017DeepRL, dulacarnold2016deep}. The most recent advancements of DRL are DDPG and soft actor-critic (SAC). DDPG is a model-free off-policy algorithm that learns continuous actions and incorporates concepts from deterministic policy gradient (DPG) and deep $Q$-network (DQN). In particular, it employs experience replay and slow-learning target networks from DQN, and it is built on DPG, which can function in continuous action spaces \cite{Lillicrap2015ContinuousCW} -- Section \ref{Sec_DDPG}. Conversely, the SAC extends the DDPG by employing a stochastic policy capable of expressing multi-modal optimum policies and entropy regularization based on the stochastic policy's entropy \cite{Haarnoja2018SoftAO}. It acts as a built-in, state-dependent exploration heuristic for the agent rather than depending on non-correlated noise processes, as in DDPG. Additionally, It incorporates two soft $Q$-networks to address the overestimation bias issue in $Q$-network-based approaches -- Section \ref{Sec_SAC}.

	\subsection{Motivation and contributions}
	
	No prior studies have explored ML techniques tailored explicitly for multiple-input multiple-output (MIMO) \bbc systems, involving a multi-antenna CE and a multi-antenna reader interacting with multiple tags (Fig.~\ref{fig_system_fig}). Importantly, our advanced DRL paradigm is also adaptable to other configurations, i.e., ambient and monostatic, offering a generalized framework that builds upon previous research \cite{Anh2019, Anh2021, Zhang2019, Liu2021}. Therefore, this research represents a pioneering study introducing DRL-based resource allocation methods specifically designed to optimize MIMO \bbc networks.
	
	With IoT networks expanding and 6G wireless technology emerging, there is a pressing need for optimized \bbc systems. Traditional methods falter in handling \bbc's complex and dynamic state and action spaces. Also, ensuring the quality of service (QoS)  for tags while maximizing long-term throughput is a significant challenge. This optimization problem is non-convex due to multi-tag interference, and its optimal solution is not amenable to widely available convex optimization tools. To address this, we develop two algorithms based on cutting-edge DRL techniques, i.e., DDPG and SAC, to offer a practical solution without relying on complex mathematical models or numerical optimization methods.  The key contributions of this paper can be summarized as follows:
	
	\begin{itemize} 
		\item This paper represents the initial effort to develop a framework that integrates advanced DRL techniques, i.e., DDPG and SAC, into the optimal design of \bbc systems to tackle high-dimensional optimization challenges.
		
		\item The focus is on maximizing the tag sum rate while satisfying the EH requirements of the tags. The CE transmit beamforming, tag reflection coefficients, and reader reception combiners are optimized to achieve this. 
		
		\item Capitalizing on DDPG and SAC, we proposed two new algorithms, namely refined-DDPG for MIMO \bbc (RDMB) and refined-SAC for MIMO \bbc (RSMB). Compared to the existing methods, our algorithms can intuitively explore and decipher the optimization problem through trial-and-error interactions with the environment. Also, they accommodate continuous state and action spaces, conferring substantial advantages over the conventional methods.
		
		\item As previously mentioned, the AO technique optimizes the transmit beamforming, reflection coefficient, and reception combiners separately, as shown in \cite{Sacarelo2021, Jia2022, Zhao2024, Diluka2024, Rezaei2024, Galappaththige2023}. In contrast, the proposed RDMB and RSMB algorithms jointly optimize these variables while progressively maximizing the sum rate by observing the reward and iteratively adjusting the algorithm parameters. However, given the ubiquity of the AO approach, we implement an AO-based benchmark for extensive comparison.
		
		\item The RDMB and RSMB algorithms feature a standard formulation and low implementation complexity, requiring no explicit model of the wireless environment or specific mathematical formulations. This makes it easily scalable to various system settings. Unlike DL-based algorithms that depend on sample labels derived from mathematically formulated algorithms, these algorithms can independently learn about and adapt to the environment.
		
		\item Particularly noteworthy is the RSMB algorithm, which outperforms the RDMB algorithm in several aspects. RSMB, using SAC, introduces an entropy regularization term into the objective function, encouraging more exploratory policy and thereby offering improved robustness and stability compared to RDMB.
		
		\item The proposed algorithms are also compared against existing ML-based methods. For this purpose, we implement  DQN, DDQN, and dueling DQN (DuelDQN) algorithms \cite{Li2017DeepRL}. Comprehensive simulations indicate that our algorithms learn incrementally from the environment and improve performance. For example, for a twelve-antenna network, RSMB leads with a \qty{26.76}{\percent} gain over DQN, followed by AO and RSMB at \qty{23.02}{\percent} and \qty{19.16}{\percent}, respectively. 
	\end{itemize}

	\textit{Notations:}  For matrix  $\mathbf{A}$, $\mathbf{A}^H$ and $\mathbf{A}^T$  represent the Hermitian conjugate transpose and transpose, respectively. $\mathbb{R}^{M\times N}$ represents $M\times N$ dimensional real matrices, and $\mathbb{C}^{M\times N}$ refers to $M\times N$ dimensional complex matrices. Euclidean norm, absolute value, and Frobenius norm are represented by $\|\cdot\|$, $|\cdot|$, and $\|\cdot\|_F$, respectively. The expectation operator is denoted by $\mathbb{E}[\cdot]$. A circularly symmetric complex Gaussian (CSCG) random vector with mean $\boldsymbol{\mu}$ and covariance matrix $\mathbf{C}$ is represented as $\sim \mathcal{C}\mathcal{M}(\boldsymbol{\mu}, \mathbf{C})$. The symbol $\qI_M$ represents an identity matrix of dimensions $M \times M$. The operators $(\cdot)^{-1}$ and $\otimes$ denote the inversion of a matrix and the Kronecker product. $\mathcal{O}$ expresses the big-O notation. Finally, $\mathcal{K} \triangleq \{1,\ldots,K\}$, $\mathcal{K}_0 \triangleq \{0, 1,\ldots,K\}$,  and $\mathcal{K}_k \triangleq \mathcal{K}\setminus\{k\}$.
	
	\begin{table}[t!]
		\centering
		\caption{Table of Notations.}
		\begin{tabular}{|L{1cm}|L{5.5cm}|}
			\hline
			\textbf{Notation} & \textbf{Description} \\
			\hline \hline
			$a$ & Action \\
			$a_{\text{NL}}$ & Nonlinear EH parameter \\
			$\alpha_k$ & Reflection coefficient of tag $T_k$ \\
			$\beta$ & Learning rate  \\
			$b_{\text{NL}}$ & Nonlinear EH parameter \\
			$D$ & Experience replay buffer capacity \\
			$\delta_d$ & Direct signal interference cancellation quality \\
			$E$ & Number of episodes \\
			$\eta$ & Power conversion efficiency \\
			$f_c$ & Carrier frequency \\
			$\mathbf{F}_0$ & Channel matrix between CE and reader \\
			$\mathbf{F}_k$ & Effective backscatter channel through tag $T_k$ \\
			$\gamma$ & Discount factor \\
			$\gamma_k$ & SINR of tag $T_k$ \\
			$\mathbf{g}_{b,k}$ & Channel between tag $T_k$ and reader \\
			$\mathbf{g}_{f,k}$ & Channel between CE and tag $T_k$ \\
			$K$ & Number of single-antenna tags \\
			$L$ & Number of experiences in mini-batch \\
			$\lambda_a$ & Actor network training decay rate \\
			$\lambda_c$ & Critic network training decay rate \\
			$M$ & Number of antennas at the CE \\
			$M_{\text{NL}}$ & Maximum harvested power \\
			$\mu_{a}$ & Actor network training learning rate \\
			$\mu_{c}$ & Critic network training learning rate \\
			$N$ & Number of antennas at the reader \\
			$\Phi(\cdot)$ & Nonlinear EH function \\
			$\pi$ & Policy \\
			$p_b$ & Activation threshold \\
			$p_b'$ & Inverse of nonlinear EH function \\
			$p_k^{\text{h}}$ & Harvested power at tag $T_k$ \\
			$P_{\text{EH}}$ & Harvested power \\
			$P_{\text{th}}$ & Threshold power \\
			$P_k^{\text{in}}$ & Incident power at tag $T_k$ \\
			$P_s$ & Maximum transmit power \\
			$P_t$ & Total transmit power of the CE \\
			$Q_{\pi}$ & State-action value function \\
			$r$ & Reward \\
			$s$ & State \\
			$\sigma^2$ & Noise power \\
			$T$ & Number of steps in each episode \\
			$\tau_{a}$ & Target actor network update learning rate \\
			$\tau_{c}$ & Target critic network update learning rate \\
			$\mathbf{u}_k$ & Receive combiner vector for tag $T_k$ \\
			$\mathbf{w}$ & Beamforming vector \\
			$\xi$ & Entropy regularization term \\ 
			$\zeta$ & Path loss exponent \\
			\hline
		\end{tabular}
		\label{tab}
	\end{table}
	
	\section{\bbc System Model}\label{General_BiBC Systemodel}
\begin{figure}[!t]\centering \vspace{-0mm}
		\def\svgwidth{220pt} 
		\fontsize{8}{8}\selectfont 
		\graphicspath{{}}
		\input{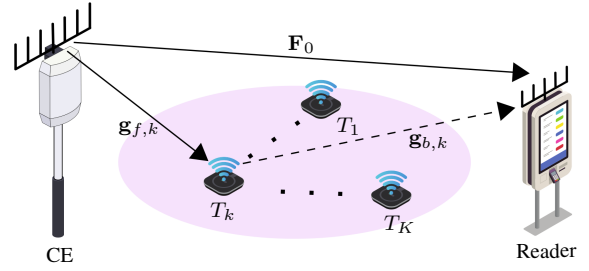} \vspace{-0mm}
		\caption{A generalized \bbc system. \vspace{-0mm}} \label{fig_system_fig}
	\end{figure}

	Fig.\ref{fig_system_fig} depicts the MIMO \bbc system consisting of an $M$-element uniform linear array (ULA) antenna CE, an $N$-element ULA antenna reader, and $K$ single-antenna tags ($T_k$ denotes the $k$-th tag for $k \in \mathcal{K}$). The ULA antennas at the CE and the reader are spaced at half-wavelength intervals \cite{Zhenyao2023}. The tags employ a power-splitting (PS) protocol for EH and data backscattering; see Section\ref{sec_EH}. This protocol ensures that the tags use part of the received CE signal for EH and reflect the remaining part to transmit data to the reader. The reader recovers the tags' data upon receiving both direct and reflected signals, although the strong direct link interference from the CE affects this process.

	\subsection{Channel Model}
	Backscatter channels differ from conventional one-way communication channels \cite{Diluka2022, Rezaei2023Coding, Zargari10320395}. The reason is that a backscatter channel is the cascade of the CE-to-tag and tag-to-reader channels.  Consequently,  it experiences double-Pathloss, leading to severe deep fades.  To model small-scale fading,   component channels are modeled as zero-mean CSCG random variables. This amounts to the Rayleigh block flat fading model with a predefined coherence time \cite{Zargari10320395, Long2020}.  
	
	During each fading block, $\mathbf{F}_0 \in \mathbb{C}^{M\times N}$, $\mathbf{g}_{f,k} \in \mathbb{C}^{M\times 1}$, and $\mathbf{g}_{b,k} \in \mathbb{C}^{N\times 1}$, respectively, represent the channels between the CE and the reader, the CE and $T_k$, and $T_k$ and the reader. These channels can be represented as $\mathbf{F}_0 = \tilde{\mathbf{F}}_0 \boldsymbol{\zeta}_{F_0}^{1/2}$ and $\mathbf{a} = \zeta_{a}^{1/2} \tilde{\mathbf{a}}$, where $\mathbf{a} \in \{\mathbf{g}_{f,k}, \mathbf{g}_{b,k} \}$. Here, $\boldsymbol{\zeta}_{F_0}$ (an $N\times N$ diagonal matrix) and $\zeta_a$ capture the large-scale Pathloss and shadowing,  which stays constant for several coherence intervals. Moreover, $\tilde{\mathbf{F}}_0 \sim \mathcal{CN}(\mathbf{0}_{M\times N}, \mathbf{I}_{M} \otimes \mathbf{I}_{N})$ and $\tilde{\mathbf{a}} \sim \mathcal{CN}(\mathbf{0}, \mathbf{I}_{A})$ account for the small-scale Rayleigh fading, where $A\in \{M,N\}$.
	
	It is assumed that the perfect channel state information (CSI) is available. A central controller, or agent (which is usually associated with the CE), can collect CSI and implement advanced DRL algorithms. The algorithm processes this state information to determine efficient actions, such as transmit beamforming, reflection coefficients, and receive combiners. The CSI can be estimated through emerging techniques such as those detailed in \cite{rezaei2023timespread, Abdallah2023, Shuo2018}. These methods include pilot-based, blind, and semi-blind approaches, utilizing algorithms like least squares, minimum mean squared error (MMSE) estimator, expectation maximization, and eigenvalue decomposition for high-precision channel estimation. However, the objective is to determine the optimal transmit beamforming, reflection coefficients, and receiver combiners by maximizing the sum rate, and leveraging recent advancements in DRL techniques for a given CSI.
	
	
	\subsection{Tag’s EH Model}\label{sec_EH}
	Because passive tags do not generate RF signals, their power consumption is extremely low, allowing them to function without batteries and rely on EH. They reflect (i.e., backscatter) external RF signals to transmit their data. PS of the incident RF signal is a critical operation at the tag \cite{Zhang2013, Azar_Hakimi}. This operation allows the tag to send data and harvest energy simultaneously. 
	
	Let the incident RF power at $T_k$ be $P_{k}^{\rm{in}}$. If $\alpha_k$ is its reflection coefficient, Tag $T_k$ uses the PS operation as follows \cite{Diluka2022, Rezaei2023Coding}:
	\begin{enumerate}
		\item It reflects $\alpha_k P_{k}^{\rm{in}}$ for data transmission,
		\item It uses the reminder, i.e., $\eta(1-\alpha_k) P_{k}^{\rm{in}}$ for EH. 
	\end{enumerate}
	The harvested power, $p_{k}^{\rm{h}}$, can be modeled as a linear or nonlinear function of $P_{k}^{\rm{in}}$. The linear model predicts harvested power at $T_k$ as $p_{k}^{\rm{h}} = \eta(1-\alpha_k) P_{k}^{\rm{in}}$, where $\eta \in (0,1]$ represents the power conversion efficiency. While the linear model is the most often utilized in the literature due to its simplicity, it disregards the nonlinear properties of actual EH circuits, such as saturation and sensitivity \cite{Wang2020WPCN}.
	
	As a result, the parametric nonlinear sigmoid EH model is frequently employed \cite{Boshkovska2015}. The harvested power at $T_k$ is calculated as $p_{k}^{\rm{h}} = \Phi((1-\alpha_k)P_{k}^{\rm{in}})$, where $\Phi(\cdot)$ is a function expressing nonlinear effects \cite[eq. (4)]{Boshkovska2015}.
	
	Despite the choice of linear or nonlinear model, the activation threshold, i.e., $p_b$, is a critical parameter. It is the minimal power required to wake up the EH circuit and is typically \qty{-20}{\dB m} for commercial passive tags \cite{Diluka2022}. To activate the tag, the harvested power must surpass the threshold, i.e., $p_{k}^{\rm{h}} \ge p_b$. In particular, $(1-\alpha_k) P_{k}^{\rm{in}} \geq  p_b'$, where $p_b' \triangleq \Phi^{-1}(p_b)$ and $ \Phi^{-1}(p_b)= b_{\rm{NL}} - \frac{1}{a_{\rm{NL}}} \ln \left( \frac{M_{\rm{NL}} - p_b}{p_b} \right)$. Parameters $ a_{\rm{NL}} $ and $ b_{\rm{NL}} $ model the circuit characteristics like capacitance and resistance. $ M_{\rm{NL}} $ denotes the maximum harvested power when the EH circuit is saturated. Parameters $a_{\rm{NL}}$, $b_{\rm{NL}}$, and $M_{\rm{NL}}$ can be derived using a curve fitting tool \cite{Boshkovska2015}. Without loss of generality, the nonlinear EH model is adopted for the rest of this paper.

	\subsection{Transmission Model}
	As mentioned before, the tags reflect the RF signal from the CE to send data to the reader.  Thus, the  CE transmits the signal $\mathbf{x} \in \mathbb{C}^{M\times 1}$ which is given by
	\begin{eqnarray}\label{eqn_BS_Tx_sig}
		\mathbf{x} = \mathbf{w}s,
	\end{eqnarray}
	where  $\mathbf{w} \in \mathbb{C}^{M\times 1}$ is the CE beamforming vector and $s=e^{-j2\pi f_c t}$ is a complex passband equivalent signal with a carrier frequency of $f_c$ and unit power, i.e., $ \mathbb{E}\{\vert s \vert^2 \}=1$ \cite{Nikos_Fasarakis}. 
	
	Assuming the propagation delay differences for all signals are negligible \cite{Liao2020}, the reader-received signal is given by
	\begin{eqnarray}\label{eqNx_reader}
		\qy = \underbrace{\qF_0^{\rm{H}} \qw s}_{\text{Direct signal interference}} + \underbrace{\sum\nolimits_{k\in \mathcal{K}} \sqrt{\alpha_k} \qF_k^{\rm{H}} \qw s c_k}_{\text{Tag signals}} + \underbrace{\qz}_{\text{AWGN}},
	\end{eqnarray}
	where  $\qz \sim \mathcal{CN}(\mathbf{0}, \sigma^2 \qI_N)$ is the additive white Gaussian noise (AWGN) at the reader, $\qF_k = \qg_{f,k} \qg_{b,k}^{\rm{H}}$ is the effective backscatter channel through $T_k$, $c_k$ is $T_k$'s data with $\mathbb{E}\{\vert c_k \vert^2 \}=1$, and $\alpha_k$ is the reflection coefficient of  $T_k$.
	
	\begin{remark}
		In \bbc, the reader eliminates the direct link signal from the CE by using a form of successive interference cancellation (SIC) before the tags' data decoding. Specifically, the direct signal in \eqref{eqNx_reader} does not carry information bits. Therefore, direct signal interference can be eliminated by estimating and subtracting the mean value of the received signal, $\mathbb{E}\{\qy\}$ \cite{Nikos_Fasarakis,biswas2021direct}.
	\end{remark}
	
	The reader then applies the receive combiner, $\qu_k\in \mathbb{C}^{N\times 1}$, to the received signal \eqref{eqNx_reader} to capture the desired signal of $T_k$ for data decoding. The post-processed signal for  decoding  $T_k$'s data is thus given as
	\begin{align}\label{eqNx_reader_post}
		y_{k} &= \underbrace{\sqrt{\delta_{d}} \qu_k^{\rm{H}} \qF_0^{\rm{H}} \qw s}_{\text{Direct interference}} + \underbrace{\sqrt{\alpha_k} \qu_k^{\rm{H}} \qF_k^{\rm{H}} \qw s c_k}_{\text{Desired signal}} \nonumber \\
		&\hspace{20mm}+\underbrace{\sum\nolimits_{i\in \mathcal{K}_k} \sqrt{\alpha_i} \qu_k^{\rm{H}} \qF_i^{\rm{H}} \qw s c_i}_{\text{Muti-tag interference}} + \underbrace{\qu_k^{\rm{H}}\qz}_{\text{AWGN}},
	\end{align}
	where $\delta_d \in [0,1]$ accounts for the degree of imperfection in the direct signal interference cancellation 
	process.  
	
	\subsection{Communication Rates}
	The reader decodes $c_k$, considering the other tag signals as interference. From \eqref{eqNx_reader_post}, $T_k$'s signal-to-interference-plus-noise ratio (SINR) at the reader can be given as
	\begin{eqnarray}\label{eqMk_SINR}
		\gamma_k = \frac{\alpha_k \vert \qu_k^{\rm{H}} \qF_k^{\rm{H}} \qw \vert^2}{\delta_d \vert \qu_k^{\rm{H}} \qF_0^{\rm{H}} \qw \vert^2 + \sum_{i\in \mathcal{K}_k} \alpha_i \vert \qu_k^{\rm{H}} \qF_i^{\rm{H}} \qw \vert^2 +\sigma^2 \Vert \qu_k \Vert^2}.
	\end{eqnarray}
	Thus, $T_k$'s rate at the reader is approximated as \cite{Long2020}:
	\begin{eqnarray}
		\mathcal{R}_{k} \approx {\rm{log}}_2(1 + \gamma_{k})\quad [\text{bps/Hz}],
	\end{eqnarray}
	Here, while the reader can partially cancel interference from the direct-link signal, represented as $\qF_0^{\rm{H}} \qw$, it cannot eliminate interference from tag reflections as they contain unknown tag data, $c_k$ for $k \in \mathcal{K}$.

	\subsection{Incident RF   Power at the tags}
	Per  Section~\ref{sec_EH}, the input power at the tags must exceed the activation threshold, typically \qty{-20}{\dB m} for commercial passive tags \cite{Diluka2022}. By considering the RF signal from the CE,  $T_k$'s input power may be expressed as 
	\begin{eqnarray}
		P_{k}^{\rm{in}}=|\q g^{\rm{H}}_{f,k}\q w|^2.
	\end{eqnarray}
	Ensuring that $P_{k}^{\rm{in}}$ exceeds the activation threshold is critical for the reliable operation of the tags \cite{Diluka2022, Rezaei2023Coding, Niu2019overview}. If the input power is below the activation threshold (see section \ref{sec_EH}), the tags may fail to activate, leading to a breakdown in data transfer processes.

	\section{Problem Formulation}
	The problem is to optimize the \bbc performance in terms of the sum rate. In particular, the objective is to maximize the tag sum rate by jointly optimizing the reader reception combiners, $\{\qu_k\}_{k\in \mathcal{K}}$, the CE transmit beamforming, $\qw$, and the tag reflection coefficients, $\{\alpha_k\}_{k\in \mathcal{K}}$ for given a particular CSI. Denote the set of these optimization variables as $\mathcal{A} = \left\{ \{\qu_k\}_{k\in \mathcal{K}}, \qw, \{\alpha_k\}_{k\in \mathcal{K}} \right\}$. The optimization problem is thus formulated as follows: 
	\begin{subequations}\label{P1_prob}
		\begin{eqnarray}
			\mathbf{P}_1:
			\underset {\mathcal{A}}{\rm{max}} && R_{\text{sum}}= \sum\nolimits_{k \in \mathcal{K}}   \mathcal{R}_{k}, \label{P1_obj} \\
			\text{s.t} && P_k^{\rm{in}} \geq P_{\rm{th}}, ~  \forall k, \label{P1_tag_EH}\\
			&& P_t \leq P_s,  \label{P1_power}\\  
			&& \Vert \mathbf{u}_k\Vert^2 = 1, ~ \forall k,  \label{P1_detector}\\
			&&  0 <\alpha_k < 1, ~  \forall  k,   \label{P1_alpha}
		\end{eqnarray}
	\end{subequations}
	where $P_{\rm{th}} = \Phi^{-1}(p_b)$. In addition, \eqref{P1_tag_EH} guarantees the minimum power requirements at the tags for EH, \eqref{P1_detector} is the normalization constraint for the reception filter at the reader, and \eqref{P1_power} limits the total transmit power of the CE with $P_s$ maximum allowable transmit power and $P_t = \Vert \mathbf{w}  \Vert^2$. Finally, \eqref{P1_alpha} is the reflection coefficient constraint at each tag. 
	
	Solving $\mathbf{P}_1$ analytically is intractable, as the problem is non-convex because the objective function in $\mathbf{P}_1$ and constraint \eqref{P1_tag_EH} both contain entangled terms involving the product of optimization variables. Although some AO-based approximation methods have been proposed to find suboptimal solutions (e.g., \cite{Sacarelo2021, Jia2022, Zhao2024, Diluka2024}), they may not work in multi-antenna multi-tag scenarios. This paper thus proposes two frameworks by leveraging recent advancements in DRL techniques, i.e., DDPG and SAC, rather than attempting to solve this highly complex optimization problem through conventional methods. Unlike traditional deep neural networks (DNNs) that require both an offline training phase and an online learning phase, our algorithms utilize each CSI to construct the state and continuously run to obtain $\mathcal{A}$ \cite{Li2017DeepRL}.
	
	Before proceeding to the proposed algorithms, it is helpful to briefly discuss the conventional optimization algorithms for $\mathbf{P}_1$, which serve as a benchmark for our solutions.
	
	\section{Conventional Optimization solutions}
	$\mathbf{P}_1$ is non-convex due to variable products in the objective and constraints. Traditional methods like gradient descent (GD) can be used to solve $\mathbf{P}_1$ \cite{boyd2004convex}. However, these solutions can be ineffective as they may get stuck in local optima and fail to provide a globally optimum solution. Thus, we resort to the AO technique, which involves splitting $\mathbf{P}_1$ into three simpler convex sub-problems: transmit beamforming, receive combiner, and tags' reflection coefficient optimization \cite{Hakimi9877898, Zargari9780612}. This approach iteratively optimizes a variable or set of variables while keeping the rest of the variables fixed until convergence is achieved \cite{boyd2004convex, bezdek2003}. We further assume the availability of perfect CSI.

	\subsection*{Sub-problem 1: Reception combiner optimization}
	Here, we design the reader receive combiners, $\{\qu_k\}_{k\in \mathcal{K}}$ for fixed $\{\qw, \{\alpha_k\}_{k\in \mathcal{K}}\}$. Since the corresponding SINR of each tag in \eqref{P1_obj} observed at the reader depends on its associate combiner vector, we maximize the sum rate by optimizing the SINR of each tag individually \cite{Azar_Hakimi}. By defining $\qh_i = \qF_i \qw$ for $i\in \mathcal{K}_0$, we can reformulate $\mathbf{P}_1$ into the following problem:
		\begin{eqnarray}\label{Pu_prob}
			\mathbf{P}_{\rm{u}}:
			\underset {\qu_k}{\rm{max}}  \frac{\qu_k^{\rm{H}} \tilde{\qh}_k \tilde{\qh}^{\rm{H}}_k \qu_k}{\qu_k^{\rm{H}} \qQ_k \qu_k}, \quad \text{s.t} \quad \Vert \mathbf{u}_k\Vert^2 = 1, ~ \forall k,
		\end{eqnarray}
	where $\tilde{\qh}_k=\sqrt{\alpha_k} \qh_k$ and $\qQ_k =  \delta_d \qh_0\qh_0^{\rm{H}} + \sum_{i\in \mathcal{K}_k} \alpha_i \qh_i \qh_i^{\rm{H}} +  \sigma^2 \qI_N$.  $\mathbf{P}_{\rm{u}}$ thus becomes a generalized Rayleigh ratio problem \cite{Stanczak2008book, Wan2016} and the  optimal received beamforming is given as \cite[\textit{Lemma 4.11}]{Stanczak2008book}
	\begin{align}\label{eqn_opt_uk}
		\qu_k^* = \frac{\qQ_k^{-1} \tilde{\qh}_k}{\|\qQ_k^{-1} \tilde{\qh}_k\|}, ~ \forall k,
	\end{align}
	which is a MMSE filter \cite{Stanczak2008book, Wan2016}.
	
	\subsection*{Sub-problem 2: Transmit beamforming optimization}
	This sub-problem optimizes $\qw$ for given $\{\{\qu_k\}_{k\in \mathcal{K}}, \{\alpha_k\}_{k\in \mathcal{K}}\}$. This is a non-convex problem as the objective function in $ \mathbf{P}_1$ is a fractional function of $\qw$, and constraint \eqref{P1_tag_EH} is non-convex.
	Accordingly, we leverage the SCA and SDR methods to solve it. SDR is widely used for beamforming optimization, enabling powerful semidefinite programming (SDP) solvers to handle large-scale problems efficiently \cite{boyd2004convex, Jia2022, Zhenyao2023}. We define the matrix $\qW \triangleq \qw\qw^{\rm{H}}$ with $\qW \succeq 0$ and $\text{Rank}(\mathbf{W}) = 1$. By doing so, the quadratic ratio function inside the $\log(\cdot)$ in the objective function is turned into its equivalent linear ratio over $\mathbf{W}$ which ends up with a concave objective function \cite{Azar_Hakimi}. By dropping rank one constraint, the relaxed convex optimization problem $\mathbf{P}_1$ is given as
	\begin{subequations}\label{Pw_prob}
		\begin{eqnarray}
			\mathbf{P}_{\rm{w}}:
			\underset {\qw}{\rm{max}} && \sum\nolimits_{k \in \mathcal{K}} \log_2(A_k(\qW)) - B_k(\qW, \qW^{(t)}), \label{Pw_obj} \\
			\text{s.t} && P_{\rm{th}} - (1-\alpha_k) \tr(\qg_{f,k}\qg^{\rm{H}}_{f,k}\qW) \leq 0, ~  \forall k, \quad \label{Pw_tag_EH}\\
			&& \tr(\qW) \leq P_s,  \label{Pw_power}\\
			&& \qW \succeq 0, 
		\end{eqnarray}
	\end{subequations}
	where $\log_2(A_k(\qW))$ and $B_k(\qW, \qW^{(t)})$ are defined in \eqref{deri_Pw} with $\qg_i = \qF_i \qu_k$ for $i\in \mathcal{K}_0$. Here, per  SCA,  $B_k(\qW, \qW^{(t)})$ is linearized using first-order Taylor series approximation near a feasible point $\qW^{(t)}$, where $\qW^{(t)}$ denotes the previous iteration values of $\qW$ \cite{Azar_Hakimi}.
	\begin{figure*}[!t]
		\begin{align}\label{deri_Pw}
			A_k(\qW)&= \delta_d\tr(\qg_0\qg_0^{\rm{H}}\qW) + \sum_{i\in\mathcal{K}}\alpha_i\tr(\qg_i\qg_i^{\rm{H}}\qW) + \sigma^2 \Vert\qu_k\Vert^2,\nonumber \\
			B_k(\qW, \qW^{(t)}) &= \log_2\left( \delta_d\tr(\qg_0\qg_0^{\rm{H}}\qW^{(t)}) + \sum_{i\in\mathcal{K}_k} \alpha_i\tr(\qg_i\qg_i^{\rm{H}}\qW^{(t)}) 
			+  \sigma^2 \Vert\qu_k\Vert^2 \right)\nonumber\\
			&\qquad \qquad \qquad+\tr\left(\frac{\delta_d\qg_0 \qg_0^{\rm{H}} + \sum_{i\in\mathcal{K}_k}\alpha_i\qg_i\qg_i^{\rm{H}} }{ \delta_d\tr(\qg_0\qg_0^{\rm{H}}\qW^{(t)}) + \sum_{i\in\mathcal{K}_k}\alpha_i\tr(\qg_i\qg_i^{\rm{H}}\qW^{(t)}) + \sigma^2 \Vert\qu_k\Vert^2} (\qW-\qW^{(t)}) \right)
		\end{align}
		\hrulefill
		\vspace{-0mm}
	\end{figure*}
	Finally, $\mathbf{P}_{\rm{w}}$ is a conventional SDP problem amenable to the  CVX tool \cite{boyd2004convex}. 
	
	\subsection*{Sub-problem 3: Reflection coefficient optimization} This  involves optimizing $\{\alpha_k\}_{k\in \mathcal{K}}$ for fixed $\{ \qw, \{\qu_k\}_{k\in \mathcal{K}}\}$. It is non-convex due to the multiple-ratio fractional objective function. To tackle this, we reformulate it by decoupling the objective's numerator and denominator. This  FP approach is motivated by Dinkelbach’s transform and is referred to as the quadratic transform because it involves quadratic terms \cite{Shen2018}. It simplifies the optimization problem by converting a non-linear fractional objective function into a more tractable form and guarantees convergence to a global optimum \cite{Shen2018}.  
	
	The resultant problem is given as
	\begin{subequations}\label{Pa_prob}
		\begin{eqnarray}
			\mathbf{P}_{\alpha}:
			\underset {\alpha_k}{\rm{max}} && \sum\nolimits_{k \in \mathcal{K}} \log_2\left(1+ 2\lambda_k \sqrt{V_k} - \lambda_k^2 S_k \right), \\
			\text{s.t} && (1-\alpha_k) \vert\qg^{\rm{H}}_{f,k}\qw \vert^2 \geq P_{\rm{th}}, ~  \forall k, \label{Pa_tag_EH}\\
			&&  0 <\alpha_k < 1, ~  \forall  k,  
		\end{eqnarray}
	\end{subequations}
	where $V_k$ and $S_k$ are the numerator and denominator of $T_k$'s SINR as functions of $\alpha_k$. Besides, $\{\lambda_k\}_{k\in \mathcal{K}}$ is the auxiliary variable introduced by the FP techniques with the optimal value of $\lambda_k^* = \sqrt{V_k}/S_k$ for $k\in \mathcal{K}$ \cite{Shen2018}. For given $\{\lambda_k\}_{k\in \mathcal{K}}$, $\mathbf{P}_{\alpha}$ is convex and can be solved via CVX Matlab \cite{Shen2018, boyd2004convex}.
	
	Algorithm \ref{alg_AO} presents the overall steps to solve $\mathbf{P}_1$ using AO. It begins by initiating random feasible solutions for $\qw$ and $\{\alpha_k\}_{k\in \mathcal{K}}$ and refines $\left\{ \{\qu_k\}_{k\in \mathcal{K}}, \qw, \{\alpha_k\}_{k\in \mathcal{K}} \right\}$ in each iteration until the normalized improvement in objective is less than $\epsilon = 10^{-3}$.

	\begin{algorithm}[!t]
		\caption{AO Algorithm}
		\begin{algorithmic}[1]
			\label{alg_AO}
			\STATE \textbf{Input}: Set the iteration counter $t = 0$, the convergence tolerance $\epsilon > 0$, initial feasible solution $\{\qw, \{\alpha_k\}_{k\in \mathcal{K}}\}$. Initialize the objective function value $F^{(0)} = 0$.  
			\WHILE{ $ \frac{F^{(t+1)} - F^{(t)}}{F^{(t+1)}} \geq \epsilon$}
			\STATE Solve \eqref{eqn_opt_uk} for optimal receive combiner, $\qu_k^{(t+1)}$.
			\STATE Solve $\mathbf{P}_{\rm{w}}$ to obtain suboptimal transmit beamforming, $\mathbf{w}^{(t+1)}$, by applying Gaussian randomization to recover the rank-one solution.
			\STATE Solve $\mathbf{P}_{\alpha}$ to obtain the suboptimal power reflection coefficients, $\alpha_k^{(t+1)}$. 
			\STATE Calculate the objective function value $F^{(t+1)}$.
			\STATE Set $t\leftarrow t+1$;
			\ENDWHILE
			\STATE \textbf{Output}: Optimal solutions $\mathcal{A}^*$.
		\end{algorithmic}
	\end{algorithm}

	\section{Configuring the DRL Environment for \bbc}
	In DRL, the agent continually interacts with the environment, executing actions and receiving immediate rewards while observing changes in the environment's state. This helps the agent to identify the best action strategy (Fig. \ref{DDPG}). We configure the DRL environment here according to the MIMO \bbc system. This is the foundation for the proposed algorithms \cite{sutton1998reinforcement}.
	
	\begin{enumerate}
		\item \textbf{Action space}: This represents an array of possible decisions. At a specific time step $t$, the agent takes an action $\qa^{(t)} \in \mathcal{A}$ based on a policy $\pi$. This action results in a transition in the environment's state, moving from the present state $\qs^{(t)}$ to the next state $\qs^{(t+1)}$. The action space of $\mathbf{P}_1$ is given by
		\begin{equation}
			\mathcal{A} = \{\qw,\alpha_k\},
		\end{equation}
		where the cardinal number of the action space is given by $D_a = 2M + K$ and $\qw = \text{Re}\{\qw\}+ \text{Im}\{\qw\}$ is separated into its real and imaginary components. It is worth mentioning that receive combiner, $\{\qu_k\}_{k\in \mathcal{K}}$, is not in the action space as it has a closed-form solution based on \eqref{eqn_opt_uk}.

		\item \textbf{State space:} This is a set of observations ($\qs^{(t)} \in S$) that describe the environment. To construct the state space for $\mathbf{P}_1$, we integrate relevant \bbc environmental information. Thus, the state space is established based on the actions/reward from $(t-1)$-th time step and all channel links and given by
		\begin{align}\label{state_space}
			\qs^{(t)} = &\bigg\{ \qa, r, P_t, P_k^{\text{in}}, \nonumber \\ 
			&\quad \text{Re}\{\mathbf{F}_0\}, \text{Re}\{\mathbf{F}_k\}, \text{Re}\{\mathbf{g}_{f,k}\}, \text{Re}\{\mathbf{g}_{b,k}\}, \nonumber \\ 
			&\quad \text{Im}\{\mathbf{F}_0\}, \text{Im}\{\mathbf{F}_k\}, \text{Im}\{\mathbf{g}_{f,k}\}, \text{Im}\{\mathbf{g}_{b,k}\} \bigg\},
		\end{align}
		where the state space dimension is represented by $D_s = 2MNK+ 2MN + 4MK+2M + 2K + 2$.  
		The states in \eqref{state_space} offer a comprehensive view of the \bbc's dynamics and interactions. The action set ($\qa$) and reward function ($r$) enable efficient decision-making based on the current state while transmit power ($P_t$) and channel links are included to reflect the \bbc environment's intricacies.

		\item \textbf{Reward function}: When an action is performed, a reward is given to the agent to evaluate how effectively the goal is achieved. The reward function takes into account both the objective function ($R_{\text{sum}}$) and the constraints of $\mathbf{P}_1$ by incorporating penalty terms, $\Omega_{\text{Pow}}$ and $\Omega_{\text{EH}}$. The reward function is given by
		\begin{equation}
			r^{(t)} =  R_{\text{sum}} - \Omega_{\text{Pow}} - \Omega_{\text{EH}},
		\end{equation}
		where 
		\begin{equation}
			\Omega_{\text{Pow}} = 
			\begin{cases} 
				0 & \text{if } \| \mathbf{w} \|^2 \leq P_s, \\
				1 & \text{otherwise}.
			\end{cases}
		\end{equation}
		\begin{equation}
			\Omega_{\text{EH}} =  \sum\nolimits_{k \in \mathcal{K}} \max\left(P_k^{\rm{in}} - \Phi^{-1}(p_b), 0\right).
		\end{equation}
		Specifically, $\Omega_{\text{Pow}}$ and $\Omega_{\text{EH}}$ ensure the transmit power and the tag's harvested power remain within the required limits.
		
		\begin{figure*}
			\centering
			\includegraphics[width=5.6in, trim={0.5cm 0.5cm 0.5cm 0.5cm}, clip]{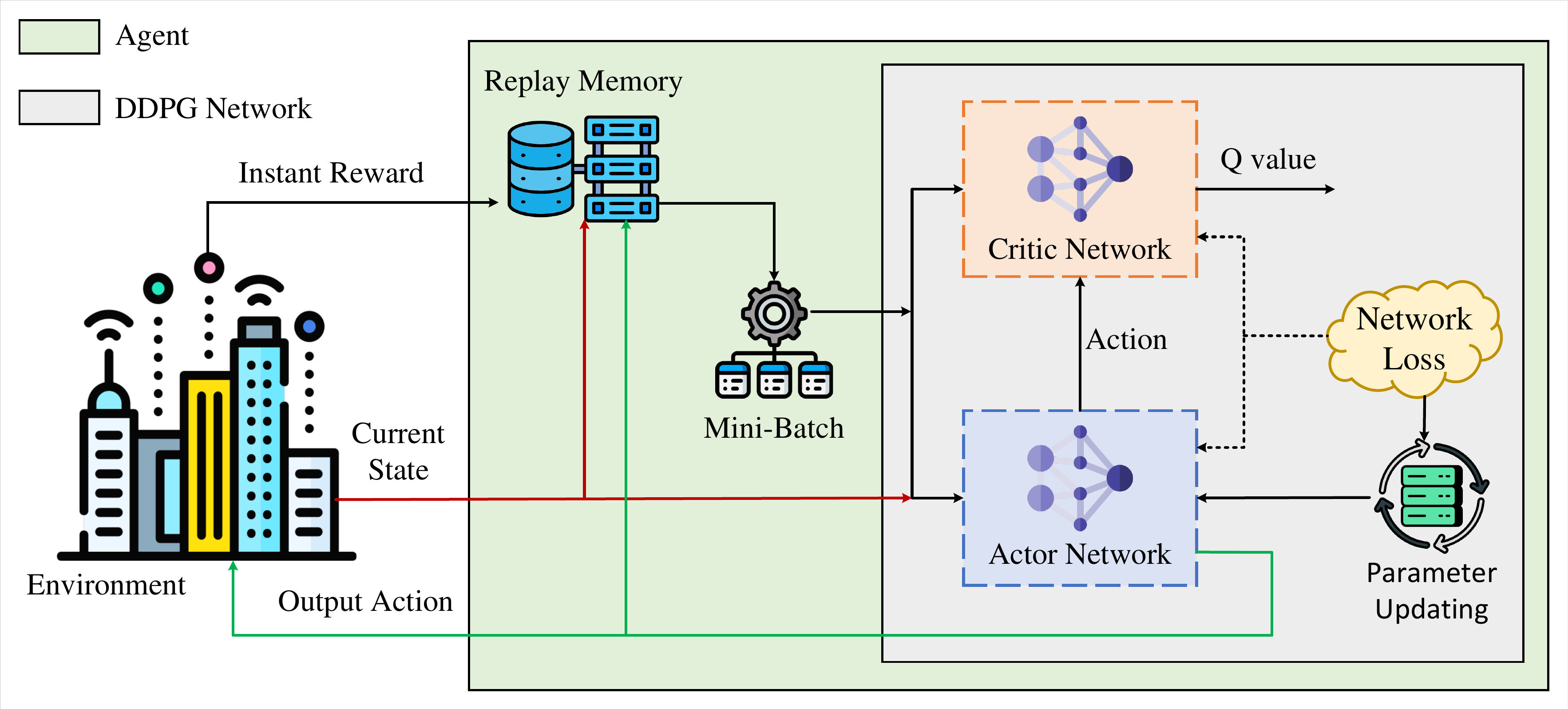}
			\caption{An illustration of deep $Q$ learning, where a double DNN approximates the optimal state-action value and $Q$-value function. \vspace{-0mm}} \label{DDPG}
		\end{figure*}
		
		\item \textbf{Policy:} The policy $\pi(\qs^{(t)}, \qa^{(t)})$ represents the probability of taking action $\qa^{(t)}$ given state $\qs^{(t)}$, ensuring that $\sum_{\qa^{(t)} \in A} \pi(\qs^{(t)}, \qa^{(t)}) = 1$.
		
		\item \textbf{State-action value function:} This represents the value of being in a particular state $s$ and performing an action $a$, defined as $Q_{\pi}(\qs^{(t)}, \qa^{(t)})$. While the reward measures the immediate return from action $a$ in state $s$, the value function estimates the potential future rewards for action $a$ in state $s$.
		
		\item \textbf{Experience replay buffer:} This stores the agent's experiences at each time step, represented as $(\qs^{(t)}, \qa^{(t)}, r^{(t+1)}, \qs^{(t+1)})$.
	\end{enumerate}

	For given $(\qs^{(t)}, \qa^{(t)}, r^{(t)})$ at time step $t$, the $Q$-value function is given by \cite{sutton1998reinforcement}:
	\begin{equation}
		Q_{\pi}(\qs^{(t)}, \qa^{(t)}) = \mathbb{E}_{\pi} \left[ R^{(t)}\; | \;\qs^{(t)} = \qs, \qa^{(t)} = \qa \right],\label{eq:Q_value_func}
	\end{equation}
	where $R^{(t)} = \sum_{\tau = 0}^{\infty} \gamma^{\tau} r(t + \tau + 1)$ and $\gamma \in (0, 1]$ denotes the discount rate. Specifically, the $Q$-value function assesses how choosing an action $\qa^{(t)}$ impacts the expected future reward under policy impacts the expected future reward under policy $\pi$. Based on the Bellman equation, the $Q$-value function can be represented as \cite{sutton1998reinforcement}:
	\begin{align}
		Q_{\pi}(\qs^{(t)}, \qa^{(t)}) &= \mathbb{E}_{\pi} \left[ r^{(t+1)}\; | \;\qs^{(t)} = \qs, \qa^{(t)} = \qa \right] 
		\nonumber \\ &\!\!\!\!\!\!\!\!\!\! +  \gamma \sum\nolimits_{\qs' \in S} P^a_{\qs\qs'} \sum\nolimits_{\qa' \in A} \pi(\qs', \qa')Q_{\pi}(\qs', \qa'),
		\label{eq:bellman_eq}
	\end{align}
	where $P^\qa_{\qs\qs'} = \Pr(\qs^{(t+1)} = \qs'\; | \;\qs^{(t)} = \qs, \qa^{(t)} = \qa)$ indicates the transition probability from state $s$ to state $\qs'$ when action $a$ is implemented \cite{sutton1998reinforcement}. The $Q$-learning algorithm aims to find the optimal policy $\pi^*$. Based on \eqref{eq:bellman_eq}, the optimal $Q$-value function associated with the optimal policy can be found as 
	\begin{align}
		Q^*(\qs^{(t)}, \qa^{(t)}) &= r^{(t+1)}(\qs^{(t)} = \qs, \qa^{(t)}, \pi = \pi^*) \nonumber \\ & + \gamma \sum\nolimits_{\qs' \in S} P^\qa_{\qs\qs'} \max_{\qa' \in A} Q^*(\qs', \qa').
		\label{eq:optimal_q_func}
	\end{align}
	By recursively solving Bellman equation \eqref{eq:optimal_q_func}, one can determine the optimal $Q^*(\qs', \qa')$ without needing precise information about the reward model or state transition model \cite{sutton1998reinforcement}. Thus, the updated $Q$-value function can be expressed as
	\begin{align}
		Q^*(\qs^{(t)}, \qa^{(t)}) &\leftarrow (1 - \beta)Q^*(\qs^{(t)}, \qa^{(t)})
		\nonumber \\ & + \beta \left( r^{(t+1)} 
		+ \gamma \max_{\qa'} Q_{\pi}(\qs^{(t+1)}, \qa') \right),
		\label{eq:update_q_func}
	\end{align}
	where $\beta$ denotes the learning rate for updating the $Q$-value function.

	Continuously updating $Q(\qs^{(t)}, \qa^{(t)})$ drives convergence towards the optimal $Q^*(\qs^{(t)}, \qa^{(t)})$ \cite{watkins1992q}. Achieving this with vast state and action spaces poses significant challenges. Function approximation techniques are often utilized to address these challenges, such as feature representation, DNNs, and methods that link value functions to state variables.
	
	The DNN employs non-linear functions to approximate the state/action-value functions and policy. Both the $Q$-value function and action are approximated by the DNN. However, DNN-based approximations lack interpretability, potentially leading DRL algorithms to attain only local optimality due to temporal correlation among states. Experience replay buffer substantially improves DRL performance by updating the DNN from a batch of randomly sampled states in the replay memory, rather than solely updating from the last state.
	
	The $Q$-value function in DRL is updated as follows:
	\begin{equation}
		Q(\qs^{(t)}, \qa^{(t)}) = Q(\qtheta\; | \;\qs^{(t)}, \qa^{(t)}).
		\label{eq:drl_q_func}
	\end{equation}
	where $\qtheta$ denotes the weight and bias parameters within the DNN. Instead of directly updating the $Q$-value function as in \eqref{eq:update_q_func}, the optimal $Q$-value function is approximated by updating $\qtheta$ through stochastic optimization algorithms, represented as
	\begin{equation}
		\qtheta^{(t+1)} \gets \qtheta^{(t)} - \mu\Delta_{\qtheta} \mathcal{L}(\qtheta),
		\label{eq:update_theta}
	\end{equation}
	where $\mu$ indicates the learning rate and $\Delta_{\qtheta}$ is the gradient of the loss function $\mathcal{L}(\qtheta)$ with respect to $\qtheta$. 
	
	The loss function measures the difference between the predicted value from the DNN and the actual target value. Although DRL involves approximating the optimal $Q$-value function, the true target value remains unknown. Two DNNs with identical architecture are designed to address this: the training DNN and the target DNN. Their value functions are respectively given by $Q(\qtheta^{\text{train}}\; | \;\qs^{(t)}, \qa^{(t)})$ and $Q(\qtheta^{\text{target}}\; | \;\qs^{(t)}, \qa^{(t)})$. The actual target value is estimated as follows:
	\begin{equation}
		y^{\text{target}} = r^{(t+1)} + \gamma \max_{\qa'} Q(\qtheta^{\text{target}}\; | \;\qs^{(t+1)}, \qa').
		\label{eq:actual_target}
	\end{equation}
	The loss function is thus defined as
	\begin{equation}
		\mathcal{L}(\qtheta) = \left(y^{\text{target}}-Q(\qtheta^{\text{train}}\; | \;\qs^{(t)}, \qa^{(t)})\right)^2.
		\label{eq:loss_func}
	\end{equation}

	\section{Refined-DDPG for MIMO \bbc} \label{Sec_DDPG}
	The proposed joint design of transmit beamforming and reflection coefficient faces a fundamental challenge.  That optimization problem has both continuous state and action spaces. We utilize the DDPG network to meet it, as depicted in Fig. \ref{DDPG}. The DDPG consists of two DNNs: (i) the actor network and  (ii) the critic network \cite{Haarnoja2018SoftAO,Huang_Chongwen}. The actor network uses the state to produce a continuous action, which is then input into the critic network along with the state. This approach approximates the action, eliminating the need to find the action that optimizes the $Q$-value function for the next state. The update rule for the critic network training is given by
	\begin{equation}
		\qtheta_{c}^{(t+1)} \gets  \qtheta_{c}^{(t)} - \mu_{c}\Delta_{\qtheta^{\text{train}}_{c}}\mathcal{L}(\qtheta^{\text{train}}_{c}), \label{eq:update-critic}
	\end{equation}
	where $\mu_{c}$ indicates the learning rate for the training critic network.  The gradient with respect to the training critic network $\qtheta^{\text{train}}_{c}$ is represented by $\Delta_{\qtheta^{\text{train}}_{c}}\mathcal{L}(\qtheta^{\text{train}}_{c})$. In addition, the loss function is defined by the mean squared error (MSE) as follows:
	\begin{equation}
		\mathcal{L}(\qtheta) = \frac{1}{L} \left\|\qr^{(t)} + \gamma q(\qtheta^{\text{target}}_{c}\; | \;\q\qs^{(t+1)}, \qa') - q(\qtheta^{\text{train}}_{c}\; | \;\q\qs^{(t)}, \q\qa^{(t)}) \right\|_2^2,
		\label{eq:theta-train-critic}
	\end{equation}
	where $(\qs, \qa, \qr, \qs')_{i=1}^{L}$  is the mini-batch of transitions sampled from the experience replay buffer with a size of $L$ and $\qa'$ denotes the action output from the target actor network. The target and training critic network parameters are symbolized by $\qtheta^{\text{target}}_{c}$ and $\qtheta^{\text{train}}_{c}$, with the target network parameters being updated like those of the training network at specific time intervals. The updating of the target network is much slower than that of the training network \cite{Haarnoja2018SoftAO,Huang_Chongwen}. The update rule for the actor network training is given by
	\begin{equation}
		\qtheta_{a}^{(t+1)} \gets  \qtheta_{a}^{(t)} - \mu_{a}\Delta_\qa q(\qtheta^{\text{target}}_{c}\; | \;\q\qs^{(t)}, \qa)\Delta_{\qtheta^{\text{train}}_{a}}\pi(\qtheta^{\text{train}}_{a}\; | \;\q\qs^{(t)}), \label{eq:update-actor}
	\end{equation}
	where $\mu_{a}$ is the learning rate for adjusting the training actor network.  $\pi(\qtheta^{\text{train}}_{a}\; | \;\q\qs^{(t)})$ is the training actor network with $\qtheta^{\text{train}}_{a}$ as the DNN parameters and $\q\qs^{(t)}$ as the given state input. The gradient of the target critic network over the action is represented by $\Delta_a q(\qtheta^{\text{target}}_{c}\; | \;\q\qs^{(t)}, \qa)$, whereas $\Delta{\qtheta^{\text{train}}_{a}}\pi(\qtheta^{\text{train}}_{a}\; | \;\q\qs^{(t)})$ represents the gradient of the training actor network over its parameter $\qtheta^{\text{train}}_{a}$. As can be inferred from \eqref{eq:update-actor}, the update process of the training actor network is guided by the target critic network through the gradient of the target critic network over the action. This ensures that the subsequent action selection follows the preferred direction of actions for optimizing the $Q$-value function \cite{Haarnoja2018SoftAO,Huang_Chongwen}.
	
	The updates on the target critic network and the target actor network are respectively given by
	\begin{equation}
		\qtheta^{\text{target}}_{c} \leftarrow \tau_{c}\qtheta^{\text{train}}_{c} + (1 - \tau_{c})\qtheta^{\text{target}}_{c},
	\end{equation}
	\begin{equation}
		\qtheta^{\text{target}}_{a} \leftarrow \tau_{a}\qtheta^{\text{train}}_{a} + (1 - \tau_{a})\qtheta^{\text{target}}_{a},
	\end{equation}
	where $\tau_{c}$ and $\tau_{a}$ are the learning rates for updating the target critic network and the target actor network, respectively.

	\begin{figure}
		\centering
		\includegraphics[width=3.2in, trim={0.5cm 0.5cm 0.5cm 0.5cm}, clip]{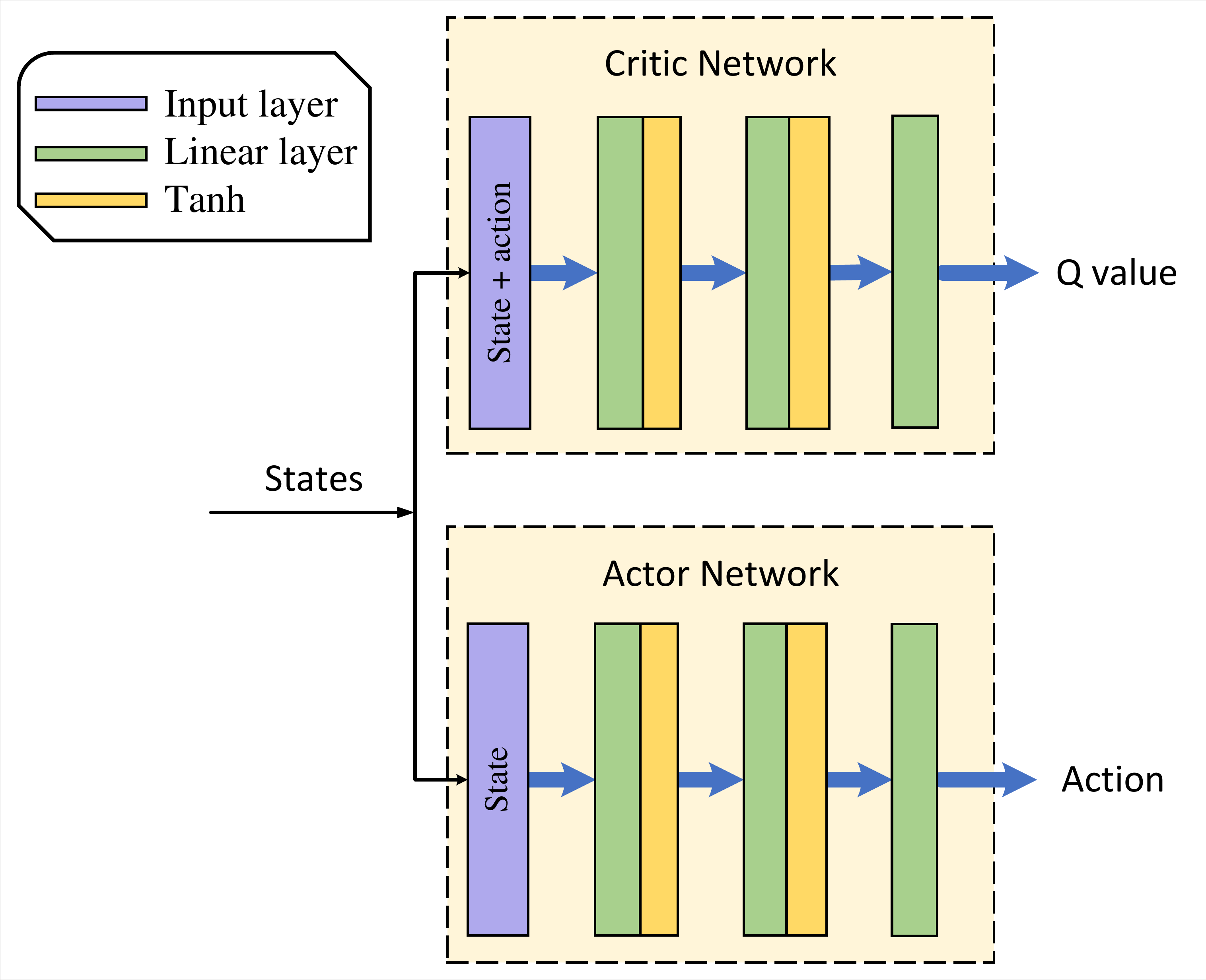}
		\caption{The DNN structure of the critic network and the actor network for the RDMB method. \vspace{-0mm}} \label{DDPG_structure}
	\end{figure}
	
	\subsection{Construction of DNN for RDMB}
	The RDMB architecture consists of fully connected layers for the critic and actor networks, as shown in Fig. \ref{DDPG_structure}.
	\begin{enumerate}
		\item  \textbf{Actor network} determines the optimal action given a particular state. The input layer corresponds to the state space. It comprises two hidden layers. The dimensions of each layer are determined by rounding up to the nearest power of $2$ that is equal to or greater than the number of state variables. The output layer corresponds to the action space with dimensions equal to the number of action variables. It is a linear layer followed by a hyperbolic tangent function (tanh) activation function \cite{goodfellow2016deep}. The output of this function is then scaled to the maximum possible action value.

		\item  \textbf{Critic network}: This evaluates the expected return, or value, of being in a specific state and taking a particular action. The input layer takes in two types of inputs: the state and the action.  The network contains two hidden layers.  The first hidden layer is a fully connected linear layer. Its dimension is set to the nearest power of two, equal to or larger than the sum of the number of state variables and action variables. After the first hidden layer processes the state input, the action input is concatenated to its output. The second hidden layer takes this concatenated vector as input.  The output layer has dimensions equal to one, representing the expected return of a given state-action pair.  
		
		\begin{algorithm}[t]
			\caption{RDMB Algorithm} 
			\label{alg:EnhancedDDPG}
			\begin{algorithmic}[1]
				\renewcommand{\algorithmicrequire}{\textbf{Input:}}
				\renewcommand{\algorithmicensure}{\textbf{Output:}}
				\newcommand{\algorithmicinitialize}{\textbf{Initialization:}}
				\REQUIRE State dimension, action space, actor and critic learning rates, $\qh_{sr}$, $\qh_{tr}$, and $h_{st}$
				\item[] \hspace{-7mm} \algorithmicinitialize\ Setup experience replay database of capacity $D$, initialize training actor network parameters $\qtheta^{(train)}_a$, set target actor network parameters $\qtheta^{(target)}_a=\qtheta^{(train)}_a$, define training critic network parameters $\qtheta^{(train)}_c$, let target critic network parameters $\qtheta^{(target)}_c=\qtheta^{(train)}_c$, generate received beamforming vector $\qw$ and reflection coefficient $\{\alpha_k\}_{k\in \mathcal{K}}$
				\FOR{ $e= 0, 1, 2, \ldots , E - 1$}
				\STATE Collect and preprocess $\qh_{sr}^{(p)}$, $\qh_{tr}^{(p)}$, $h_{st}^{(p)}$ for the $e$-{th} iteration to generate the initial state $s^{(0)}$
				\FOR{$t=0, 1, 2, \ldots , T - 1$}
				\STATE Derive action $\qa(t) = \{\qw^{(t)}, \{\alpha_k\}_{k\in \mathcal{K}}^{(t)}\} = \pi(\qtheta^{(train)}_a)$ from the actor network
				\STATE Solve \eqref{eqn_opt_uk} for optimal receive combiner, $\{\qu\}_{k\in \mathcal{K}}^{(t)}$
				\STATE \quad  Determine new $\q\qs^{(t+1)}$ using action $\q\qa^{(t)}$
				\STATE \quad Record immediate reward $\qr^{(t+1)}$
				\STATE \quad Preserve the experience $(\q\qs^{(t)}, \q\qa^{(t)}, \qr^{(t+1)}, \q\qs^{(t+1)})$\\ \quad in the replay database
				\STATE Extract the $Q$ value function as $Q = q(\qtheta^{(train)}_c\; | \;\q\qs^{(t)}, \q\qa^{(t)})$ from the critic network
				\STATE Randomly draw mini-batches of experiences of size $L$ from the replay database 
				\STATE Develop training critic network loss function $\mathcal{L}(\qtheta^{(train)}_c)$ as indicated in \eqref{eq:theta-train-critic}
				\STATE Apply SGD on training critic network to obtain $\Delta_{\qtheta^{\text{train}}_{c}}\mathcal{L}(\qtheta^{\text{train}}_{c})$
				\STATE Apply SGD  on target critic network to obtain $\Delta_a q(\qtheta^{\text{target}}_{c}\; | \;\q\qs^{(t)}, \qa)$
				\STATE Apply SGD on training actor network to obtain $\Delta_{\qtheta^{(train)}_a} \pi(\qtheta^{(train)}_a\; | \;\q\qs^{(t)})$
				\STATE Update the training critic parameters $\qtheta^{(train)}_c$
				\STATE Load DNN with input as $\q\qs^{(t+1)}$
				\ENDFOR
				\ENDFOR
				\RETURN Optimal action $\qa = \{\qw^*, \{\alpha_k\}_{k\in \mathcal{K}}^*, \{\qu\}_{k\in \mathcal{K}}^*\}$
			\end{algorithmic}
		\end{algorithm}

	\end{enumerate}

	In both networks, the tanh function is used as the activation function for the hidden layers to effectively handle negative inputs \cite{goodfellow2016deep}. The training process for both the critic and actor networks employs the Adam optimizer, which uses adaptive learning rates \cite{Lillicrap2015ContinuousCW,goodfellow2016deep}. The learning rates are updated at each time step according to $\mu^{(t)}_c = \lambda_c\mu^{(t-1)}_c$ and $\mu^{(t)}_a = \lambda_a\mu^{(t-1)}_a$, where $\mu^{(t)}_c$ and $\mu^{(t)}_a$ represent the learning rates for the critic and actor networks at time step $t$, respectively. The parameters $\lambda_c$ and $\lambda_a$ are the decay rates used to gradually reduce the learning rates over time, promoting convergence during training \cite{Haarnoja2018SoftAO,Huang_Chongwen,sutton1998reinforcement}.

	A detailed description of the proposed RDMB is presented in Algorithm \ref{alg:EnhancedDDPG}. It assumes a central controller or agent that can instantaneously gather channel information, i.e., $\{\mathbf{F}_0,  \mathbf{g}_{f,k}, \mathbf{g}_{b,k}\}$. At each time step $t$, the agent employs the present channel information and the action executed previously, i.e., $(\qw^{(t-1)}, \{\alpha_k\}_{k\in \mathcal{K}}^{(t-1)})$, to formulate the current state $\qs^{(t)}$. Furthermore, we initialize $\qw$ with the identity matrix and set $\{\alpha_k\}_{k\in \mathcal{K}}$ to a value of $0.5$. The algorithm operates over $E$ episodes, with each episode iterating $T$ steps. The optimal $\qw^*$ and $\{\alpha_k\}_{k\in \mathcal{K}}^*$ are determined by the action that yields the highest immediate reward.

	\begin{figure}
		\centering
		\includegraphics[width=3.5in, trim={0.5cm 0.5cm 0.5cm 0.5cm}, clip]{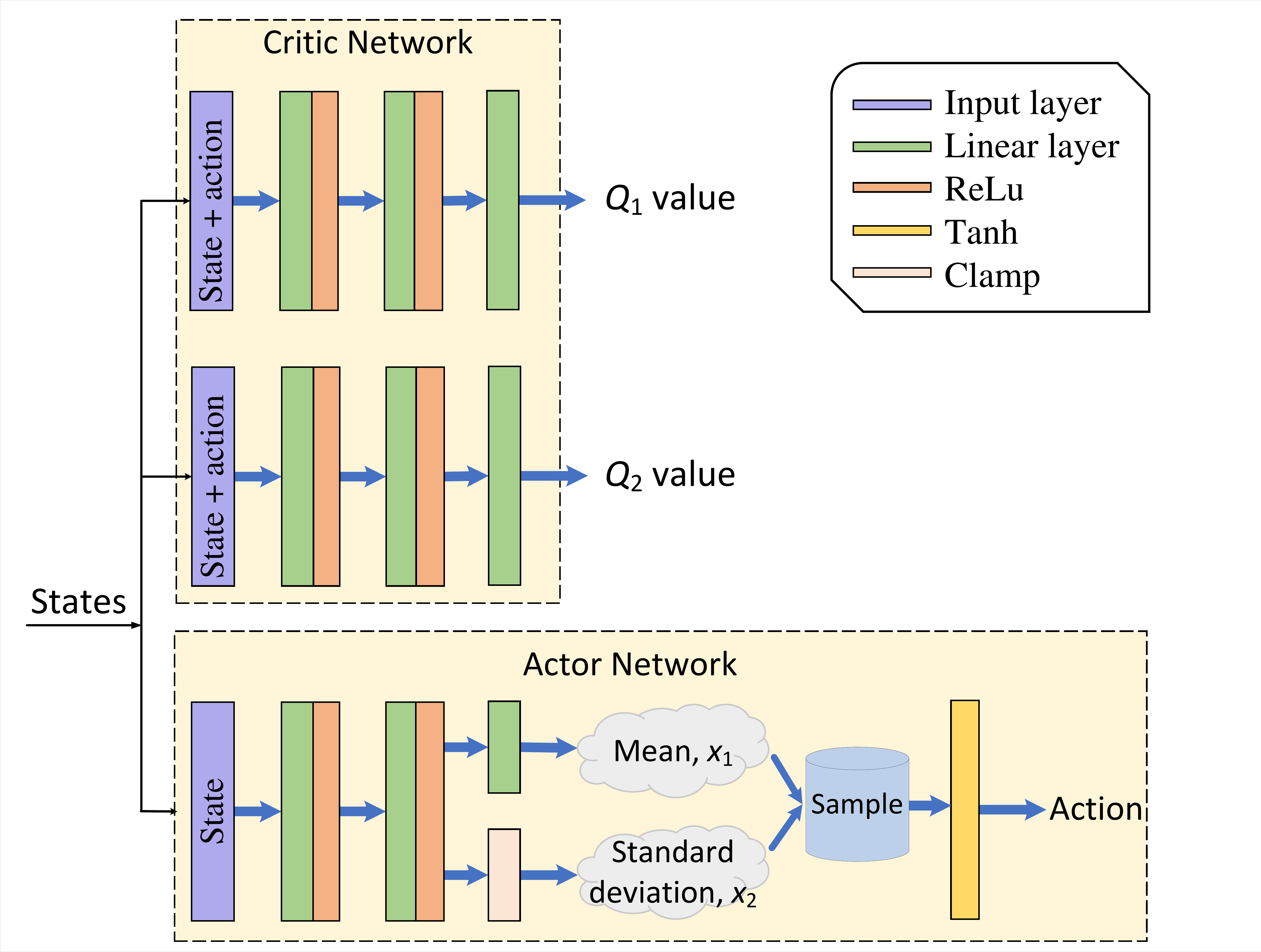}
		\caption{The DNN structure of the critic network and the actor network for the RSMB method. \vspace{-0mm}} \label{SAC_structure}
	\end{figure}
	
	\section{Refined-SAC for MIMO \bbc}\label{Sec_SAC}
	The SAC method is a state-of-the-art maximum entropy DRL algorithm that optimizes the trade-off between maximizing the cumulative reward and maintaining a high entropy policy to encourage adequate exploration \cite{Haarnoja2018SoftAO}. The proposed RSMB algorithm employs the policy iteration approach, alternating between policy evaluation and policy improvement steps \cite{saglam2023deep}. This technique is off-policy, making it efficient and scalable. It allows learning from experiences stored in a replay buffer. This algorithm primarily involves three networks: two $Q$-networks and one policy actor network. The two $Q$-networks, i.e., $Q_{\qtheta_{1}}$ and $Q_{\qtheta_{2}}$, mitigate the overestimation of $Q$-value estimates, a common issue in DRL algorithms \cite{sutton1998reinforcement,Haarnoja2018SoftAO}.
	
	The $Q$-networks utilize states provided by the environment, and actions generated by the actor network to derive $Q$-value estimates. Denote the actor network as $\pi_{\qpsi}$, where $\qpsi$ represents the policy parameters. In the RSMB algorithm, the $Q$-networks are trained jointly with the following rule:
	\begin{equation}
		\hat{\qy} =  \qr + \gamma \min_{i=1,2} Q_{\qtheta_{i}'} (\qs', \qa')\big|_{\qa' \sim\pi_{\qpsi}(\cdot \vert \qs')} - \xi \log(\qa' \; | \; \qs'), \label{eq:24}
	\end{equation}
	where $(\qs, \qa, \qr, \qs')$ indicates a transition obtained from the replay buffer. The $\qtheta_{i}$ denotes the parameters of the $i$-th $Q$-network, $\xi$ is the entropy regularization term or temperature, and $\mu$ is the learning rate. Accordingly, The loss function is defined by the MSE between the target value $y^{\text{target}}$ and the current $Q$-value estimate as follows:
	\begin{equation}
		\mathcal{L}(\qtheta_{i}) = \frac{1}{L} \left\|\hat{\qy}  - Q_{\qtheta_{i}} (\qs, \qa) \right\|_2^{2}.\label{eq:26}
	\end{equation}
	The update on the parameters of the $i$-th $Q$-network is given by
	\begin{equation}
		\qtheta_{i}^{(t+1)}  \gets \qtheta_{i}^{(t)} - \mu \nabla_{\qtheta_{i}} \mathcal{L}(\qtheta_{i}).\label{eq:25}
	\end{equation}
	Similarly, the policy network utilizes state vectors from the environment to produce action vectors. The SAC framework calculates the policy network loss through:
	\begin{equation}
		\mathcal{L}(\qpsi)  = \frac{1}{L} \sum_{i=1}^{L} \xi \log \pi_{\qpsi}(\hat{\qa}_{i} \; | \; \qs_{i}) - \min_{j=1,2} Q_{\qtheta_{j}} (\qs_{i}, \hat{\qa}_{i})\bigg|_{\hat{\qa}\sim\pi_{\qpsi}(\cdot \vert \qs)}.\label{eq:27}
	\end{equation}
	Subsequently, the policy gradient $\nabla_{\qpsi} \mathcal{L}(\qpsi)$ is computed via the stochastic policy gradient algorithm  and applied to update the parameters using gradient ascent:
	\begin{equation}
		\qpsi^{(t+1)} \gets \qpsi^{(t)} + \mu \nabla_{\qpsi} \mathcal{L}(\qpsi).\label{eq:28}
	\end{equation}
	In particular, the degree of exploration within the system is primarily regulated by the entropy regularization term, $\xi$. Higher values of $\xi$ encourage more exploration. Although a deterministic policy-based DRL algorithm can be utilized, it requires additional noise for efficient exploration. Conversely, the SAC approach's entropy regularization feature utilizes the current policy knowledge, enhancing its ability to handle complicated such as received beamforming and reflection coefficient design.

	\begin{algorithm}[t]
		\caption{RSMB Algorithm} 
		\label{alg:SAC}
		\begin{algorithmic}[1]
			\renewcommand{\algorithmicrequire}{\textbf{Input:}}
			\renewcommand{\algorithmicensure}{\textbf{Output:}}
			\newcommand{\algorithmicinitialize}{\textbf{Initialization:}}
			\REQUIRE State dimension, action space, actor and critic learning rates, $\xi$, $\qh_{sr}$, $\qh_{tr}$, and $h_{st}$
			\item[] \hspace{-7mm} \algorithmicinitialize\ Setup experience replay database  of capacity $D$, initialize critic network $Q_{\qtheta_1}$ and $Q_{\qtheta_2}$, actor network $\pi_{\qpsi}$, the parameters for training, generate received beamforming vector $\qw$ and reflection coefficient $\{\alpha_k\}_{k\in \mathcal{K}}$
			\FOR{ $e= 0, 1, 2, \ldots , E - 1$}
			\STATE Preprocess the state and action for the $e$-th episode
			\FOR{$t=0, 1, 2, \ldots , T - 1$}
			\STATE Derive action $\q\qa^{(t)}$ from the actor network by sampling from the policy $\pi_{\qpsi}$
			\STATE Solve \eqref{eqn_opt_uk} for optimal receive combiner, $\{\qu\}_{k\in \mathcal{K}}^{(t)}$
			\STATE \quad Detect new state $\q\qs^{(t+1)}$ using action $\q\qa^{(t)}$ 
			\STATE \quad Compute immediate reward $\qr^{(t+1)}$  
			\STATE \quad Preserve the experience $(\q\qs^{(t)}, \q\qa^{(t)}, \qr^{(t+1)}, \q\qs^{(t+1)})$ \\ \quad in the replay buffer
			\STATE Sample a mini-batch from the replay buffer
			\STATE Compute the target $Q$-value for each tuple in the mini-batch using \eqref{eq:24} and the critic target network
			\STATE Update the critic networks by minimizing the loss function $\mathcal{L}(\qtheta_{i})$ as defined in \eqref{eq:26} 
			\STATE Update the $i$-th $Q$-network parameter based on \eqref{eq:25}
			\STATE Compute the policy loss $\mathcal{L}(\qpsi)$ as defined in \eqref{eq:27} with the help of the minimum predicted $Q$-value of the two critic networks
			\STATE Optimize the actor by performing gradient ascent as per \eqref{eq:28} on the expected return
			\STATE Adjust $\xi$ by optimizing it towards maximizing entropy
			\ENDFOR
			\ENDFOR
			\RETURN Optimal action $\qa = \{\qw^*, \{\alpha_k\}_{k\in \mathcal{K}}^*, \{\qu\}_{k\in \mathcal{K}}^*\}$
		\end{algorithmic}
	\end{algorithm}

	\subsection{Construction of DNN for RSMB}
	Fig. \ref{SAC_structure} demonstrates the critic and actor DNN structures used for the RSMB method.  
	
	\begin{enumerate}
		\item \textbf{Actor network}: This generates a distribution over actions for a given state. The input layer corresponds to the state space. This network has two fully connected linear layers as its hidden layers. The dimensions of these layers are calculated as the nearest power of two larger than or equal to the number of state variables. The output layer comprises two parts. One part outputs the mean of the action distribution, and the other outputs the log standard deviation of the action distribution. Both are fully connected linear layers with dimensions equal to the number of action variables.  To ensure numerical stability, the outputted standard deviation is clamped between predefined minimum and maximum values.

		\item \textbf{Critic network}: This predicts the $Q$-value for a given state-action pair. The input layer takes in both the state and the action.  This network has two sets of hidden layers, effectively forming two separate sub-networks. Each sub-network has two fully connected linear layers. The dimensions of these layers are determined by the nearest power of $2$ that is larger than or equal to the sum of the number of state variables and the number of action variables. Each sub-network has an output layer with a single unit, which outputs the estimated $Q$-value for the given state-action pair. The critic network's final output is the minimum of these two $Q$-value estimates to provide a conservative estimate.
	\end{enumerate}

	\begin{table*}[t]
		\centering
		\caption{Comparison of DRL Algorithms}
		\begin{tabular}{|l|p{4.5cm}|p{3.5cm}|p{4.5cm}|}
			\hline
			\textbf{Algorithm} & \textbf{Key Feature} & \textbf{Application} & \textbf{Advantage} \\
			\hline \hline
			SAC & Entropy-based DRL & Continuous action spaces & Balances exploration and exploitation efficiently \\
			\hline
			DDPG & Actor-critic method with off-policy learning & High-dimensional, continuous action spaces & Stable and robust to hyper-parameter settings \\
			\hline
			DQN & Value-based with DL & Discrete action spaces & Simple implementation, good in stable environments \\
			\hline
			DDQN & Reduces overestimation by decoupling selection and evaluation & Discrete action spaces & Improves over DQN in stability \\
			\hline
			Dueling DDQN & Separates state-value and advantage-value estimation & Discrete action spaces & Provides finer control over $Q$-value estimation \\
			\hline
		\end{tabular}
		\label{table:algorithm_comparison}
	\end{table*}
	
	All networks utilize the rectified linear unit  (ReLU) as the activation function for hidden layers and the Adam optimizer for optimization with adaptive learning rates. Model parameters are updated using experiences sampled from the replay buffer. The critic network minimizes the disparity between its $Q$-value estimates and target $Q$-values derived from the actor's policy, while the actor network maximizes the expected $Q$-value estimated by the critic and policy entropy. The entropy coefficient ($\xi$) adjusts dynamically during training to balance exploration and exploitation. Critic target network parameters are periodically soft-updated from the critic network. The SAC method is outlined in Algorithm \ref{alg:SAC}.

	\begin{algorithm}[t]
		\caption{Unified Deep $Q$-Learning Algorithm}
		\label{alg:improved_unified_dql}
		\begin{algorithmic}[1]
			\renewcommand{\algorithmicrequire}{\textbf{Input:}}
			\renewcommand{\algorithmicensure}{\textbf{Output:}}
			\newcommand{\algorithmicinitialize}{\textbf{Initialization:}}
			\REQUIRE Environment with states $S$, actions $\mathcal{A}$, and reward function $R$
			\algorithmicinitialize{}
			Setup replay memory to capacity $D$, action-value function $Q$ with random weights $\qtheta$, target action-value function $\hat{Q}$ with weights $\qtheta^- = \qtheta$, exploration rate $\epsilon$, network architecture (DQN, DDQN, Dueling)
			\FOR{ $e= 0, 1, 2, \ldots , E - 1$}
			\STATE Preprocess the state and action for the $e$-th episode
			\FOR{ $t=0, 1, 2, \ldots , T - 1$}
			\STATE Select action $a_t$ based on $\epsilon$-greedy policy  from $Q(\qs_t, \qa; \qtheta)$ 
			\STATE Execute action $\qa_t$ in emulator and observe reward $r_t$ and next state $\qs_{t+1}$  
			\STATE Store transition $(\qs_{t}, \qa_t, r_t, \qs_{t+1})$ in replay memory
			\STATE Sample random minibatch $(\qs_{t}, \qa_t, r_t, \qs_{t+1})$ from replay memory 
			\IF{DDQN}
			\STATE Set \\$\hat{y}_t = r_t + \gamma \hat{Q}(\qs_{t+1}, \arg\max_{\qa'} Q(\qs_{t+1}, \qa'; \qtheta); \qtheta^-)$  
			\ELSIF{Dueling}
			\STATE Compute $V(\qs_{t})$ and $A(\qs_{t}, \qa_t)$ using separate streams; combine to form $Q(\qs_{t}, \qa_t)$
			\STATE Set $\hat{y}_t = r_t + \gamma \big(V(\qs_{t+1}) + A(\qs_{t+1}, \qa')$ \\ \hspace{40mm} $- \frac{1}{|A|} \sum_{\qa'} A(\qs_{t+1}, \qa') \big)$ 
			\ELSE
			\STATE Set $\hat{y}_t = r_t + \gamma \max_{\qa'} \hat{Q}(\qs_{t+1}, \qa'; \qtheta^-)$ 
			\ENDIF
			\STATE Perform GD on $(\hat{y}_t - Q(\qs_{t}, \qa_t; \qtheta))^2$ with respect to $\qtheta$
			\STATE Every $C$ step, update $\hat{Q}$ with weights from $Q$  
			\STATE Reduce $\epsilon$ gradually (e.g., $\epsilon \times$ decay factor) 
			\ENDFOR
			\STATE Evaluate policy $\pi$
			\ENDFOR
			\ENSURE Optimal action $\qa = \{\qw^*, \{\alpha_k\}_{k\in \mathcal{K}}^*\}$
		\end{algorithmic}
	\end{algorithm}

	\section{Deep $Q$-Benchmarks}
	This section presents DQN-based benchmarks, including DQN, DDQN, and DuelDQN, to evaluate the performance of the proposed algorithms.

	DQN and its variants are inherently designed for discrete action spaces \cite{Li2017DeepRL}. Applying DQN to problems with continuous action spaces requires discretizing these spaces, which enables the use of DQN but at the cost of lower sum rate performance. This can result in less effective learning and diminished performance than actor-critic architectures \cite{Li2017DeepRL}.
	
	The trade-off here involves a balance between performance and complexity \cite{sutton1998reinforcement}. DQN methodologies, while more straightforward and easier to implement due to their reliance on discrete action spaces, tend to suffer from lower performance in terms of sum rate and learning efficiency when applied to continuous problems. This is because discretization reduces the granularity with which actions can be taken, leading to potential suboptimal decision-making \cite{mnih2015human}. On the other hand, actor-critic architectures, although more complex, can directly handle continuous action spaces. This allows for more precise action selection and potentially better overall performance, especially in environments where fine-grained control is crucial. Therefore, the choice between these approaches depends on the specific requirements of the problem, such as the acceptable level of performance and the available computational resources \cite{van2016deep,Haarnoja2018SoftAO}.
	
	In DQN methodologies, we need to have a discrete action space \cite{van2016deep}. This process is generally outlined as follows:
	\begin{enumerate}
		\item \textbf{Transmit beamforming:} The selection of the beamforming vector is executed from a predetermined codebook. This codebook contains an array of potential beamforming configurations \cite{Mismar8938771}. It is assumed that the CE employs analog-only beamforming vectors. The beamforming weights for a vector $\qw$ are effectuated utilizing constant-modulus phase shifters. Additionally, it is assumed that the beamforming vector is derived from a beamsteering-based beamforming codebook $\mathcal{W}$ with a cardinality $|\mathcal{W}| := L_{\text{CE}}$. Thus, the $l$-th element within this codebook is defined as
		\begin{equation}
			\!\!\!\! \qw_l = \frac{1}{\sqrt{M}} \left[ 1, e^{jkd \cos(\qtheta_l)}, \ldots, e^{jkd(M-1) \cos(\qtheta_l)} \right]^T\!\!, \!
		\end{equation}
		where $d$ and $k$ denote the antenna spacing and the wavenumber, respectively, while $\qtheta_l$ denotes the steering angle.  
		
		\item \textbf{Transmit power:} The transmit power is divided into $L_P=5$ equally sized intervals, ranging from a minimum of $0$ to a maximum power of $P_{s}$. This division creates a set $\mathcal{P}$, which consists of values starting from $0.1$ up to $P_{s}$, evenly distributed across $L_P=5$  points.
		
		\item \textbf{Reflection coefficients:} The reflection coefficients $\{\alpha_k\}_{k \in \mathcal{K}}$ are quantized into $L_{\text{EH}} = 5$ discrete levels. The set $\mathcal{E}$ is constructed using evenly spaced values between $\frac{1}{L_{\text{EH}} - 1}$ and $1 - \frac{1}{L_{\text{EH}} - 1}$, resulting in a total of $L_{\text{EH}}$ points.
		
	\end{enumerate}
	
	\subsection{Deep $Q$-Networks}
	We present an integrated approach combining DL strategies with $Q$-learning. At the heart of the DQN framework lies a complex DNN architecture \cite{sutton1998reinforcement,van2016deep}. It is crafted to provide precise estimates of the $Q$ function, which is expressed as $Q(\qs, \qa) = \mathbb{E}[r_{t+1} + \gamma \max_{\qa'} Q(\qs', \qa') \mid \qs, \qa]$, where it predicts the expected rewards for taking action $\qa$ in state $\qs$ and following the optimal policy thereafter \cite{van2016deep}.
	
	The DNN is trained with a variant of the Bellman equation as its loss function, denoted as $L(\qtheta) = \mathbb{E}[(\hat{y}_t - Q(\qs, \qa; \qtheta))^2]$ \cite{sutton1998reinforcement,van2016deep}. Here, $\hat{y}_t =r_{t+1} + \gamma \max_{\qa'} Q(\qs', \qa'; \qtheta) $ denotes the target $Q$-value, with $\qtheta$ representing the unified set of parameters for action selection and evaluation. Additionally, the DQN methodology enhances learning efficiency and stability by implementing an experience replay mechanism. This mechanism involves storing and randomly sampling experiences, which minimizes the correlation between consecutive learning samples \cite{lin1992self}.

	\subsection{Double Deep $Q$-Networks}
	
	The DDQN represents an evolution of the DQN model, specifically aimed at mitigating the overestimation of $Q$-values inherent in the original DQN approach \cite{van2016deep}. DDQN achieves this by decoupling the action selection and action evaluation processes \cite{van2016deep}. The revised equation for calculating $Q$-values in DDQN is given by
	\begin{equation}
		Q(\qs, \qa) = r_{t+1} + \gamma Q(\qs', \underset{\qa'}{\mathrm{argmax}}\ Q(\qs', \qa'; \qtheta), \qtheta^-).
	\end{equation}
	In this setup, the decision regarding which action $\qa'$ to take is made using the primary network, characterized by $\qtheta$. However, this action's $Q$-value is evaluated using a separate target network with parameters $\qtheta^-$. This bifurcation significantly contributes to a reduction in the overestimation of $Q$-values \cite{van2016deep}.
	
	Moreover, DDQN introduces an improved methodology for updating the target network, opting for a gradual or “soft” update mechanism rather than a direct transfer of parameters from the primary to the target network \cite{Li2017DeepRL}. This method allows for a more gradual alignment of $\qtheta^-$ with $\qtheta$, thereby ensuring a smoother update process for the target network and ultimately leading to a more stable learning environment \cite{van2016deep}.
	
	\subsection{Dueling Deep $Q$-Networks}
	DuelDQN introduces a significant modification to traditional DQN  by separately evaluating the state value function $V(\qs)$ and the advantage function $A(\qs, \qa)$ \cite{wang2016dueling}. Advantage functions, $A(\qs, \qa)= Q(\qs, \qa) - V(\qs)$, are designed to distinguish between the value derived from being in a certain state $s$ and the value derived from executing a certain action $a$ in that state, with the formal definition being $A(\qs, \qa) = Q(\qs, \qa) - V(\qs)$ \cite{wang2016dueling}. After the initial feature extraction within the network, DuelDQN proceeds to calculate the $Q$-value as follows:
	\begin{equation}
		Q(\qs, \qa) = V(\qs) + \left( A(\qs, \qa) - \frac{1}{|\mathcal{A}|} \sum\nolimits_{\qa'} A(\qs, \qa') \right).
	\end{equation}
	This highlights the architecture's capability to independently learn the intrinsic value of each state value function and the relative advantage of each action within that state. The latter is normalized by subtracting the average advantage, stabilizing the training phase \cite{wang2016dueling}.

	\begin{table}[t]
		\centering
		\caption{Simulation Hyper-Parameter}
		\label{table:1}
		\begin{tabular}{|l|c|}
			\hline
			\textbf{Parameter} & \textbf{Default Value} \\
			\hline \hline
			Maximum transmit power  ($P_s$) & $40$ dBm \\
			\hline
			Carrier frequency ($f_c$) & \qty{3}{\GHz} \\
			\hline
			Minimum required EH ($\Pth$) & $-20$ dBm  \\
			\hline
			The number of antennas at the CE/reader ($M/N$) & 12 \\
			\hline
			Experience replay buffer capacity ($D$) & $100000$ \\
			\hline
			Number of experiences in the mini-batch ($L$) & $32$  \\
			\hline
			Number of steps in each episode ($T$) & $10$  \\
			\hline
			Number of episodes ($E$) & $5000$  \\
			\hline
			Future reward discount factor ($\gamma$) & 0.99  \\
			\hline
			Actor network training learning rate ($\mu_{a}$) & $10^{-3}$  \\
			\hline
			Critic network training learning rate ($\mu_{c}$) & $10^{-3}$  \\
			\hline
			Target actor network update learning rate ($\tau_{a}$) & $10^{-3}$  \\
			\hline
			Target critic network update learning rate ($\tau_{c}$) & $10^{-3}$ \\
			\hline
			Actor network training decay rate ($\lambda_a$) & $10^{-5}$ \\
			\hline
			Critic network training decay rate ($\lambda_c$) & $10^{-5}$  \\
			\hline
			Initial coefficient of entropy ($\xi$) & $0.2$ \\
			\hline
			SAC log standard deviation clipping & $(-20, 2)$  \\
			\hline
			Entropy target & $-$action dimension  \\
			\hline
			Non-linear EH parameters ($\{a_{\rm{NL}}, b_{\rm{NL}}\}$) & $\{6400, 0.003\}$ \\
			\hline
		\end{tabular}
	\end{table}
	
	\begin{table*}[ht]
		\centering
		\caption{Overall computational complexity of different methods.}
		\begin{tabular}{|m{1.5cm}|m{10cm}|}
			\hline
			\textbf{Algorithm} & \textbf{ Complexity} \\ 
			\hline \hline
			AO & $\mathcal{O}\left( \mathcal{O}(KN^2 + N^3) +  T \left( \mathcal{O}(I_1(2KM^2 + M + N)^{4.5})  + \mathcal{O}(I_2K^{3.5}\log (1/\epsilon)) \right) \right)$ \\ 
			\hline
			RDMB & $\mathcal{O} \left(E T B \left(2 \sum_{i=0}^{2} l_{a_i} l_{a_{i+1}} + 2 \sum_{i=0}^{2} l_{c_i} l_{c_{i+1}}\right)  + |D|  \sum_{i=0}^{2} l_{a_i} l_{a_{i+1}}\right)$ \\ 
			\hline
			RSMB & $\mathcal{O} \left(E T B \left(2 \sum_{i=0}^{2} l_{a_i} l_{a_{i+1}} + 4 \sum_{i=0}^{2} l_{c_i} l_{c_{i+1}}\right)  + |D|  \sum_{i=0}^{2} l_{a_i} l_{a_{i+1}}\right)$ \\ 
			\hline
			DQN & $\mathcal{O} \left( E T B  \left( 2  \sum_{i=0}^{2} l_i  l_{i+1} \right) + |D|  \left( \sum_{i=0}^{2} l_{i} l_{i+1}  \right) \right)$ \\ 
			\hline
			DDQN & $\mathcal{O}\left(  E T B  \left(4 \sum_{i=1}^{2} l_i  l_{i+1}\right)+ |D|   \left(\sum_{i=0}^{2} l_{i} l_{i+1}  \right) \right)$  \\ 
			\hline
			DuelDQN & $\mathcal{O} \left( E T B  \left(2\left(\sum_{i=0}^{1} l_i  l_{i+1} + l_v+l_aa \right)\right) + |D|  \left(2 \left(\sum_{i=0}^{1} l_i  l_{i+1} + l_v+l_aa \right) \right) \right)$  \\ 
			\hline
		\end{tabular}
		\label{tab:complexity_comparison}
	\end{table*}
	
	\section{Computational Complexity Analysis}
	Table \ref{tab:complexity_comparison} compares the total computational complexity of the proposed and benchmark algorithms, where $T$ is the number of time steps, $E$ denotes episodes, $B$ indicates batch size, $|D|$ is the size of the replay buffer, and $l_{a_i}$ and $l_{c_i}$ are layer dimensions for actor and critic networks, respectively.
	
	The computational complexities for the AO algorithm arise from three subproblems \cite{bezdek2003}:
	\begin{itemize}
		\item The first involving matrix operations with a complexity of $\mathcal{O}(KN^2 + N^3)$;
		\item The second, a standard SDP problem with $\mathcal{O}(I_1(2KM^2 + M + N)^{4.5})$;
		\item The third uses the interior point method with the complexity of $\mathcal{O}(I_2K^{3.5}\log(1/\epsilon))$.
	\end{itemize}
	For DRL methods:
	\begin{itemize}
		\item The online application phase primarily involves the actor network's execution in the proposed RDMB and RSMB algorithms, contributing to a moderate complexity \cite{Haarnoja2018SoftAO,Huang_Chongwen}.
		\item The training phase for RDMB and RSMB escalates the computational load due to the simultaneous training of actor and critic networks \cite{Haarnoja2018SoftAO,Huang_Chongwen}. 
		\item For DQN, the complexity in both the online and training phases arises from executing a single network \cite{sutton1998reinforcement}.
		\item DDQN introduces a second network during the training phase to reduce overestimation by decoupling the selection from the action evaluation, doubling the training complexity compared to DQN \cite{van2016deep}.
		\item DuelDQN modifies the network architecture to separately estimate the state value and the advantages for each action, which adds complexity due to the additional computations for these separate components \cite{wang2016dueling}.
	\end{itemize}
	The overall computational complexity for training is structured around the number of neurons in each network layer, the number of episodes, steps per episode, batch size, and the capacity of the replay memory \cite{sutton1998reinforcement,Haarnoja2018SoftAO,Huang_Chongwen}.

	In particular, RDMB and RSMB, suitable for continuous action spaces, balance complexity with performance, with RSMB offering better robustness due to entropy regularization. DQN, practical for discrete action spaces, has lower computational complexity at the cost of a lower sum rate, with DDQN and DuelDQN enhancing stability and learning efficiency by addressing overestimation bias and separating state value from the advantage function, respectively. These complexities highlight the trade-off between computational demands and performance.
	
	\section{Simulation Results}\label{sim_results}
	To evaluate the proposed resource allocation methods, we consider a MIMO \bbc network where the CE and the reader are positioned at coordinates $(3, 0, 0)\qty{}{m}$ and $(0, 8, 0)\qty{}{m}$, respectively. We assume that the tags are uniformly distributed within a circle of radius $r = \qty{2}{m}$, centered at $(3, 8, 0)\qty{}{m}$. The Pathloss is given by $L(d) = C_0 \left(\frac{d}{d_0}\right)^{-\zeta}$, where $C_0=\qty{-30}{\dB}$ represents the Pathloss at the reference distance $d_0 = \qty{1}{m}$, and $d$ denotes the distance between the transceivers. We set $\zeta=\num{2}$ and use the Rayleigh fading model for channels. Unless otherwise specified, all simulation hyperparameters are provided in Table \ref{table:1}. The noise power is given by $\sigma_u^2 = -147 +10\log_{10}(\text{BW})+\text{NF}$, where bandwidth (BW) and noise figure (NF)  are $10^{6}\qty{}{\Hz}$ and $\qty{10}{\dB}$, respectively. The training process encompasses a maximum of $\num{5000}$ episodes, with each episode comprising $\num{10}$ time slots, resulting in a cumulative total of $\num{50000}$ training steps. 
	
	The proposed RSMB and RDMB algorithms are compared against two benchmarks: 1) the AO method and 2)  deep Q-network methods of DQN, DDQN, and DuelDQN. The minimum exploration rate for the agent, the initial exploration rate for the agent, the exploration decay rate for the agent, and number of steps between target network updates for deep Q-networks are set to $\num{0.01}$, $\num{1.0}$, $\num{0.995}$, and $\num{1000}$, respectively.

\begin{figure}[t]
		\centering
		\includegraphics[width=3.2in]{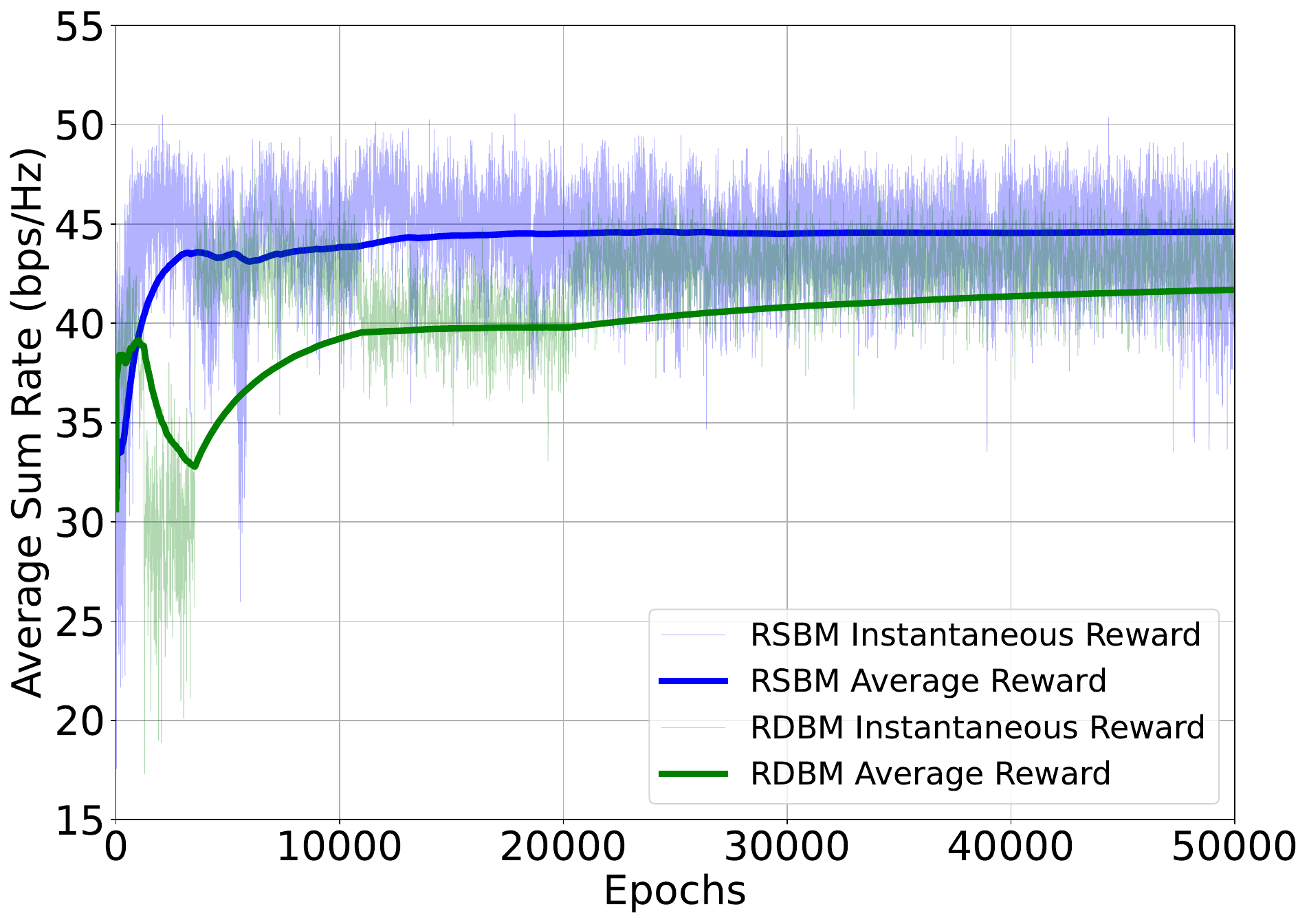}
		\caption{Average and instantaneous sum rate versus epoch for proposed schemes when $N=M=10$ and $P_s=40$ dBm. \vspace{-0mm}} \label{fig_1}
	\end{figure}

	Fig. \ref{fig_1} illustrates the RSMB algorithm's enhanced capability in achieving higher average sum rates over epochs compared to the RDMB approach. The smoother convergence pattern observed with RSMB can be attributed to SAC's entropy-regularization feature, which promotes exploring a broad range of actions through a stochastic policy framework. This feature is especially beneficial in complex scenarios characterized by large, high-dimensional state spaces, such as those encountered in MIMO systems. Conversely, the RDMB algorithm exhibits more variability in its learning curve, likely due to its actor-critic structure's proneness to accumulating errors in the $Q$-function estimation process. Additionally, RDMB's performance is sensitive to the tuning of hyperparameters, including learning rates and the size of the replay buffer, which can lead to fluctuations in its effectiveness. Unlike RSMB's stochastic policy approach, RDMB's reliance on deterministic policies may limit its ability to explore the action space thoroughly. Despite these challenges, RSMB maintains a high average sum rate, demonstrating its superior ability to balance exploration and exploitation of well-known profitable strategies, making it a highly suitable algorithm for real-world applications where environments are neither stationary nor entirely predictable.
	
	\begin{figure}[t]
		\centering
		\includegraphics[width=3.25in]{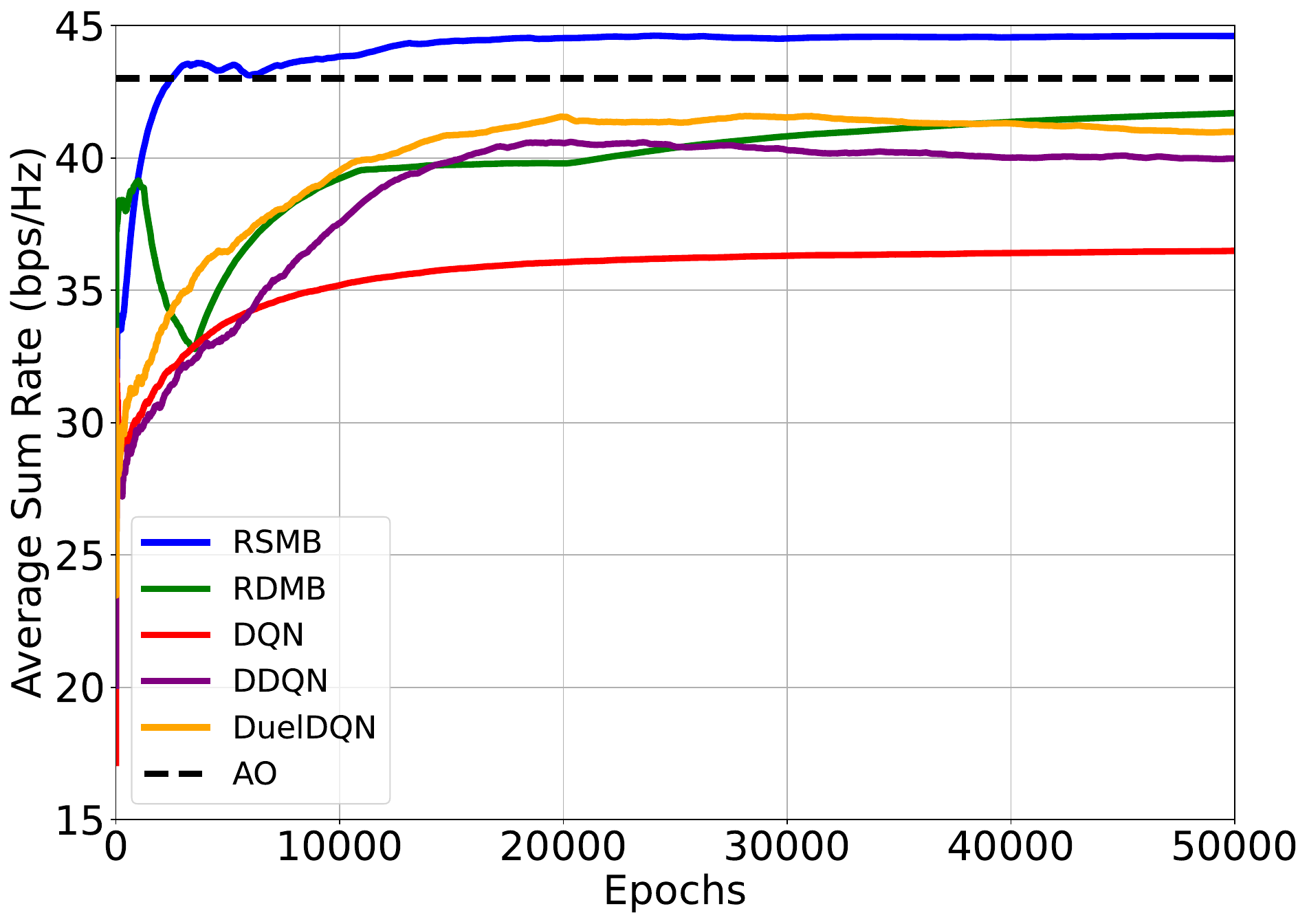}
		\caption{Average sum rate versus epoch for different schemes when $N=M=10$, $P_s=40$ dBm. \vspace{-0mm}} \label{fig_2}
	\end{figure}
	
	\begin{figure}[t]
		\centering
		\includegraphics[width=3.25in]{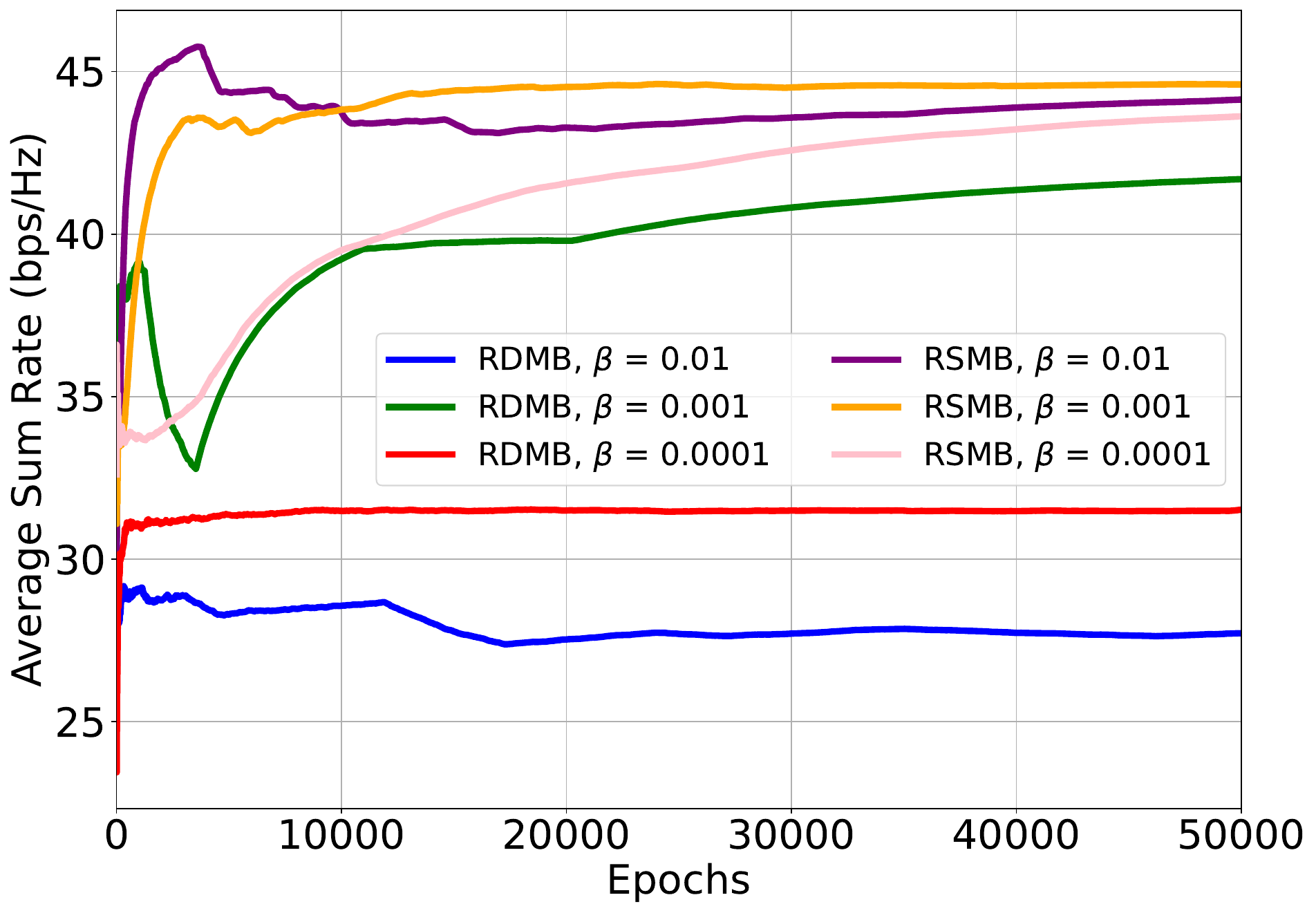}
		\caption{Average sum rate versus epoch for proposed schemes under different learning rates when $N=M=10$ and $P_s=40$ dBm. \vspace{-0mm}} \label{fig_3}
	\end{figure}

	Fig. \ref{fig_2} assesses the performance of various algorithms when applied to the MIMO \bbc network, focusing on their adaptability to continuous, high-dimensional action spaces. Among the algorithms evaluated, the RSMB stands out for its ability to effectively navigate the intricate optimization landscape of MIMO systems. RSMB's design combines off-policy learning with an entropy-driven exploration mechanism, enabling it to converge toward the AO method. Assuming perfect CSI, the AO benchmark provides a valuable metric for assessing the practical utility of DRL algorithms in resource allocation scenarios. Also, RSMB's implementation of twin-delayed updates can reduce the overestimation bias often observed in the value network, allowing for a more accurate approximation of the AO benchmark. This feature contrasts with RDMB, which is prone to overestimating value functions through its single $Q$-value estimation mechanism, potentially leading to the adoption of suboptimal policies.
	
	In comparison, value-based DRL algorithms such as DQN, DDQN, and DuelDQN, typically more suited to environments with discrete action spaces, fall short in this setting. Their slower convergence is mainly due to the challenges associated with discretizing the continuous action space and managing the curse of dimensionality inherent in MIMO. Conversely, policy gradient techniques like SAC and DDPG are naturally more compatible with the continuous nature of the action spaces in MIMO networks, as they learn policies directly. 
	
	Fig.  \ref{fig_3} delves into how variations in the learning rate influence the performance of the RSMB and RDMB algorithms. The analysis reveals RDMB's acute responsiveness to changes in the learning rate, indicating that its policy updates might become overly cautious or precipitate training instability at different rates. A reduced learning rate could stabilize the training by moderating the pace of updates; however, this adjustment risks decelerating the overall learning trajectory. Conversely, RSMB exhibits a remarkable resilience to fluctuations in learning rate, a characteristic attributed to its dual-network structure. This relative insensitivity of RSMB to learning rate modifications is further enhanced by its entropy regularization feature, which systematically encourages a thorough investigation of the action space. 
	
	\begin{figure}[t]
		\centering
		\includegraphics[width=3.15in]{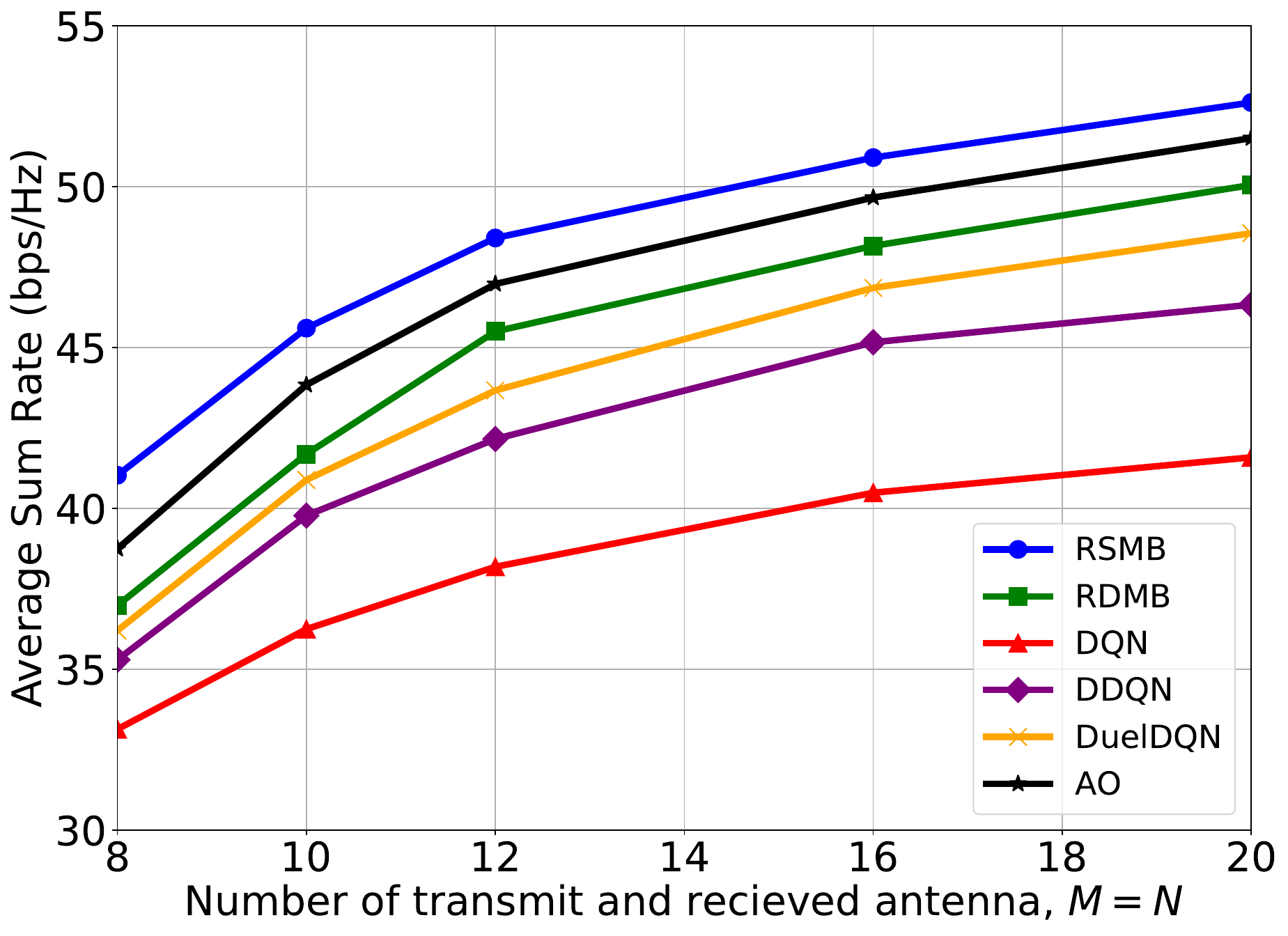}
		\caption{Average sum rate versus number of antennae at the reader for different schemes when $N=M=10$ and $P_s=40$ dBm. \vspace{-0mm}} \label{fig_4}
	\end{figure}
	
	\begin{figure}[t]
		\centering
		\includegraphics[width=3.2in]{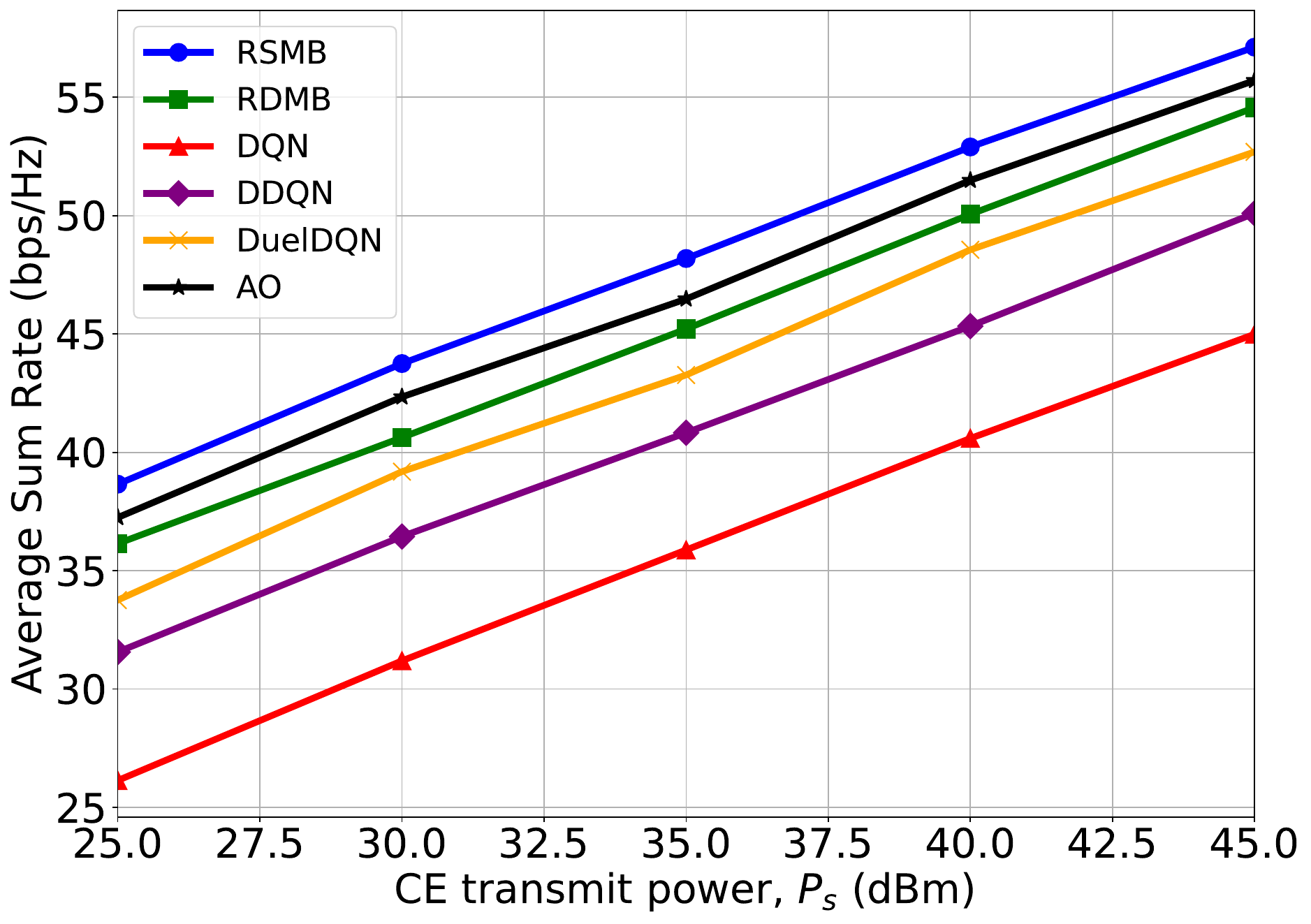}
		\caption{Average sum rate versus CE transmit power for different schemes when $N=M=20$. \vspace{-0mm}} \label{fig_5}
	\end{figure}
	
	Fig. \ref{fig_4} explores the impact of increasing the number of antennas on the performance of various algorithms. This observation corroborates the theoretical expectation that additional antennas contribute to greater spatial diversity and multiplexing gains. Notably, the RSMB algorithm exhibits a more pronounced performance improvement with the antenna array's expansion, indicating its superior capability to exploit the augmented spatial opportunities for optimizing MIMO \bbc systems. This advantage is likely rooted in SAC's proficiency in managing high-dimensional continuous action spaces. In comparison, other algorithms, such as the RDMB, also benefit from adding antennas, but to a lesser extent. This figure suggests that RSMB is well-positioned to take advantage of such advancements, making it a compelling choice for adaptive and scalable resource allocation in MIMO \bbc.
	
	When $N=M=\num{12}$, it is observed that the RSMB scheme exhibits the highest performance gain of \qty{26.76}{\percent}, compared to DQN. AO and RDMB follow this with gains of \qty{23.02}{\percent} and \qty{19.16}{\percent}, respectively. In contrast, DDQN and DuelDQN report more modest improvements of \qty{10.40}{\percent} and \qty{14.36}{\percent} when compared to DQN. Further analysis, with RDMB serving as the baseline, reveals RSMB's superiority with a \qty{6.38}{\percent} increase in performance, while AO demonstrates a \qty{3.24}{\percent} improvement. This detailed comparison underscores the effectiveness of RSMB in enhancing the MIMO \bbc system.
	
	\begin{figure}[t]
		\centering
		\includegraphics[width=3.35in]{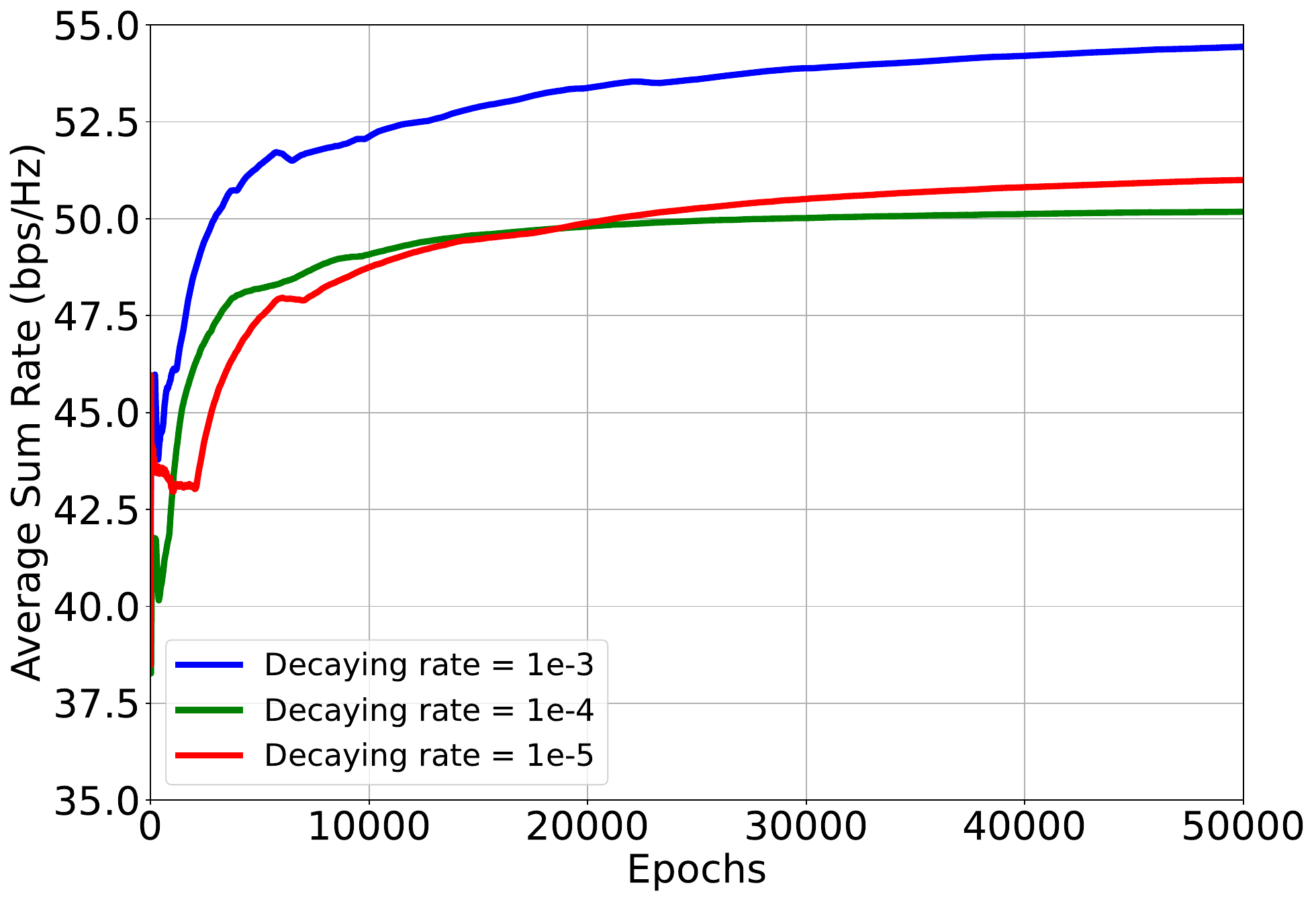}
		\caption{Average sum rate versus epochs for DDPG under different decaying rates when $N=M=20$ and $P_s=40$ dBm. \vspace{-0mm}} \label{fig_6}
	\end{figure}
	
	\begin{figure}[t]
		\centering
		\includegraphics[width=3.3in]{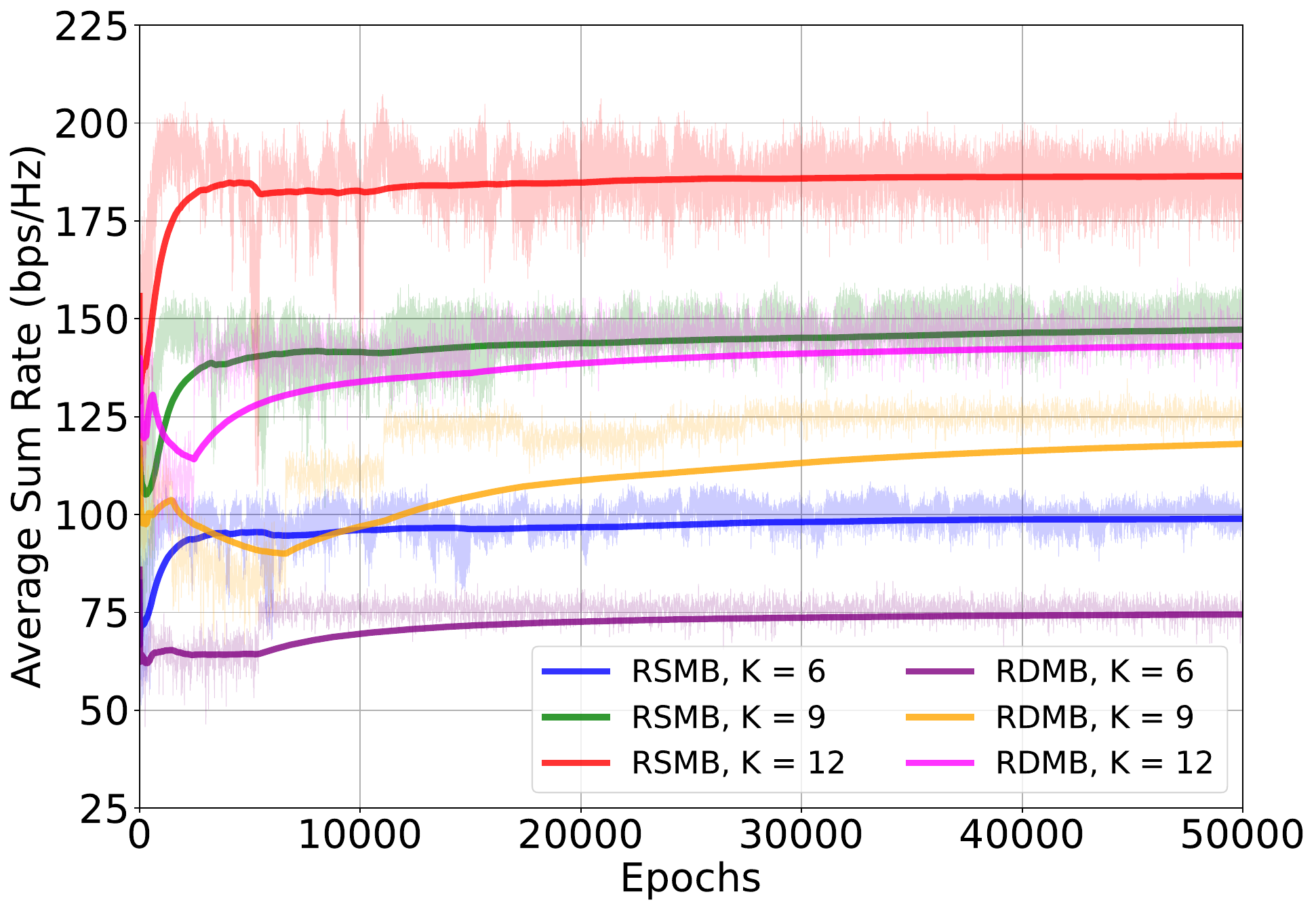}
		\caption{Average sum rate versus epochs for proposed schemes for different numbers of users when $N=M=20$ and $P_s=40$ dBm. \vspace{-0mm}} \label{fig_7}
	\end{figure}
	
	Fig. \ref{fig_5} compares the performance of various algorithms with the CE transmit power ($P_s$). In particular, RSMB stands out for its exceptional performance, showcasing an advanced approach to power management. This adaptability is essential for developing wireless devices that are power-efficient, cost-effective, and long-lasting.  By reducing energy use, RSMB enhances network efficiency and supports many applications and services that demand high performance. 
	
	Fig. \ref{fig_6} delves into how different decay rates affect the learning trajectory of the RDMB algorithm. The decay rate dictates how the learning rate decreases over time. A gradual decay rate, depicted by the $1e^{-5}$ curve, enables the algorithm to refine its policy further after initial convergence. This gradual approach is advantageous in complex scenarios, where finding the optimal policy requires navigating through a landscape fraught with local optima. Conversely, a decay rate of $1e^{-3}$ leads to a rapid early performance boost but soon reaches a plateau. This early plateau suggests the algorithm may converge prematurely to suboptimal policies.
	
	In Fig. \ref{fig_7}, we observe the effect of varying the number of backscatter devices on the performance of RSMB and RDMB algorithms. A key observation is that both algorithms benefit from the increased number of devices, which likely corresponds to more opportunities for constructive multi-user interference and potential capacity enhancement. The RSMB algorithm, however, seems to leverage these opportunities more efficiently, as indicated by its superior performance at higher $K$ values. RSMB’s ability to maintain a stable performance despite the increased state space suggests that its policy updates are robust to the curse of dimensionality.

	\section{Conclusion}\label{section_Conclusion}
	This paper addressed the intricate challenge of maximizing throughput in MIMO \bbc systems through the joint optimization of CE transmit beamforming, tag reflection coefficients, and reader reception combiners while adhering to the EH requirements of the tag. Two advanced DRL techniques, DDPG and SAC, designed for continuous state and action spaces, are exploited to propose two new algorithms (i.e., RDMB and RSMB). These algorithms employ iterative trial-and-error interactions with the environment to refine their strategies. RSMB incorporates an entropy regularization term that encourages exploration, enhancing its robustness and stability over RDMB. Our comprehensive simulation results demonstrate that RSMB outperforms AO and DQN methodologies, and in turn, AO shows superior performance over RDMB. These findings highlight the effectiveness of the advanced SAC method, as it incrementally improves and surpasses established benchmark techniques. Looking ahead, our research will pivot towards exploring multi-agent DRL frameworks to manage scenarios involving multiple readers and tags.
	
	\bibliographystyle{IEEEtran}
	\bibliography{IEEEabrv,Ref_new}

\end{document}